\documentclass{aa}

\usepackage{graphicx}
\usepackage{txfonts}

\newcommand{\Mpc}{$h^{-1}$\thinspace Mpc}

\newcommand{\etal}{{\rm et al.~}}
\def\apj{ApJ}
\def\apjl{ApJL}
\def\apjs{ApJS}

\def\aap{A\&A}

\begin{document}   

\title{Superclusters of galaxies from the 2dF redshift survey. \\
I. The catalogue} 

\author{ J. Einasto\inst{1} \and M. Einasto\inst{1}\and
E. Tago\inst{1} \and E. Saar\inst{1} \and G. H\"utsi\inst{1} \and
M. J\~oeveer\inst{1} \and L. J. Liivam\"agi\inst{1} \and
I. Suhhonenko\inst{1} \and J. Jaaniste\inst{2} \and
P. Hein\"am\"aki\inst{3} \and V. M\"uller\inst{4} \and
A. Knebe\inst{4} \and D. Tucker\inst{5} }

\institute{Tartu Observatory, EE-61602 T\~oravere, Estonia
\and 
Estonian University of Life Sciences
\and
Tuorla Observatory, V\"ais\"al\"antie 20, Piikki\"o, Finland 
\and
Astrophysical Institute Potsdam, An der Sternwarte 16,
D-14482 Potsdam, Germany
\and
 Fermi National Accelerator Laboratory, MS 127, PO Box 500, Batavia,
IL 60510, USA
}

\date{ Received 2006; accepted} 

\authorrunning{J. Einasto et al.}

\titlerunning{2dFGRS supercluster catalogue}

\offprints{J. Einasto }

\abstract {We use the 2dF Galaxy Redshift Survey data to compile
  catalogues of superclusters for the Northern and Southern regions of
  the 2dFGRS, altogether 543 superclusters at redshifts $0.009 \leq z \leq
  0.2$. 
  We analyse methods of compiling supercluster catalogues
  and use results of the Millennium Simulation to investigate possible
  selection effects and errors. We find that the most effective
  method is the density field method using smoothing with an Epanechnikov
  kernel of radius 8~\Mpc.  We derive positions of the highest
  luminosity density peaks and find the most luminous cluster in the
  vicinity of the peak, this cluster is considered as the main cluster
  and its brightest galaxy the main galaxy of the supercluster. In
  catalogues we give equatorial coordinates and distances of
  superclusters as determined by positions of their main clusters.  We
  also calculate the expected total luminosities of the superclusters. 


}

\maketitle

\section{Introduction}

It is presently well established that galaxies form various systems
from groups and clusters to superclusters. Galaxy systems are not
located in space randomly: groups and clusters are mostly aligned to
chains (filaments), and the space between groups is populated with
galaxies along the chain.  The largest non-percolating galaxy systems
are superclusters of galaxies which contain clusters and groups of
galaxies with their surrounding galaxy filaments.

Superclusters of galaxies have been used for a wide range of studies.
Superclusters are produced by large-scale density perturbations which
evolve very slowly. Thus the distribution of superclusters contains
information on the large-scale initial density field, and their
properties can be used as a cosmological probe to discriminate
between different cosmological models. The internal structure of
superclusters conserves information on the galaxy formation and
evolution on medium scales.  Properties of galaxies and groups in
various supercluster environments can be used to study the evolution of
galaxies on small scales. Superclusters are massive density
enhancements and thus great gravitational attractors which distort the
background radiation, yielding information on the gravitation field
through the CMB distortion via the Sunyaev-Zeldovich effect, which can be
detected using new satellites, such as  PLANCK.

Early studies of superclusters of galaxies were reviewed by Oort
(\cite{oort83}) and Bahcall (\cite{bahcall88}). These studies were
based on observational data about galaxies, as well as on data about
nearby groups and clusters of galaxies.  Classical, relatively deep
all-sky supercluster catalogues were constructed using the Abell
(\cite{abell}) and Abell \etal (\cite{aco}) cluster catalogues by
Zucca \etal (\cite{z93}), Einasto \etal (\cite{e1994}, \cite{e1997},
\cite{e2001}) and Kalinkov \& Kuneva (\cite{kk95}).

The modern era of the study of various systems of galaxies began when
new galaxy redshift surveys began to be published.  The first of
such surveys was the Las Campanas Galaxy Redshift Survey, followed by
the 2 degree Field Galaxy Redshift Survey (2dFGRS) and the Sloan
Digital Sky Survey (SDSS).  These surveys cover large regions of the sky
and are rather deep allowing to investigation of the distribution of
galaxies and systems of galaxies out to fairly large distances from us.
Catalogues of superclusters were compiled on the basis of these new
surveys by Einasto et al. (\cite{e03a}, \cite{e03b}, hereafter E03a
and E03b), Basilakos (\cite{bas03}), Erdogdu et al. (\cite{erd04}) and
Porter \& Raychaudhury (\cite{pr05}).  These observational studies
have been complemented by the analysis of the evolution of superclusters
and the supercluster-void network (Shandarin, Sheth, \& Sahni
\cite{shandarin04}, Einasto et al. \cite{ein05b}, Wray, Bahcall et
al. \cite{wray06}).

\begin{table}[ht]
      \caption[]{The 2dF samples used}
         \label{Tab1}
\begin{center}
         \begin{tabular}{cccrcc}
            \hline\hline
            \noalign{\smallskip}
          Sample  & RA & DEC & $N_{gal}$ &
        $N_{scl}$ & $V$ \\
             & deg & deg &    &  & \\
            \noalign{\smallskip}
            \hline
            \noalign{\smallskip}

2dFN  & 147.5~~223 & -6.3~~+2.3  & 78067   & 229 & 12.42 \\
2dFS  & 325~~55  & -36.0~~-23.5 & 106328   & 314 & 18.71 \\

            \noalign{\smallskip}
            \hline
         \end{tabular}\\
\end{center}
\end{table}

So far the attention of astronomers has been focused either on
small compact systems, such as groups and clusters, or on very large
systems -- rich superclusters of galaxies. Modern redshift surveys
make it possible to investigate not only these classical systems of
galaxies, but also galaxy systems of intermediate sizes and richness
classes of various sizes, from poor galaxy filaments in large cosmic
voids to rich superclusters. In the present series of papers we shall
discuss the richest of these systems -- superclusters of galaxies.  We
shall use the term ''supercluster'' for galaxy systems larger than
groups and clusters which have a certain minimal mean overdensity of
the smoothed luminosity density field but are still non-percolating.
They form intermediate-scale galaxy systems between groups and poor
filaments and the whole cosmic web.

The main goal of this paper is to compile a new catalogue of
superclusters using the 2dFGRS. To get a representative statistical
sample we include in our catalogue superclusters of all richness
classes, starting from poor superclusters of the Local Supercluster
class, and ending with very rich superclusters of the Shapley
Supercluster class.  In the compilation of a statistically homogeneous
and complete sample of superclusters we make use of the possibility to
recover the true expected total luminosity of galaxy systems, using
weights to compensate the absence of galaxies from the sample which
are too faint to fall within the observational window of the survey.
The use of weights has some uncertainties, so we have to investigate
errors and possible biases of our procedure to recover the total
luminosity of superclusters.  To investigate selection effects and
biases we shall investigate properties of simulated superclusters
based on the catalogue of galaxies of the Millennium Simulation of the
evolution of the structure of the Universe by Springel et
al. (\cite{springel05}).  In an accompanying paper we shall
investigate properties of superclusters (Einasto et al. \cite{e06},
hereafter Paper II).  A similar study using the SDSS is in
preparation.

The paper is composed as follows. In the next Section we shall
describe the observational and model data used. In Section 3 we shall
discuss superclusters in the cosmic web.  In Section 4 we shall
discuss selection effects and their influence on supercluster
catalogues. This section is based on simulated superclusters using the
Millennium Simulation.  Sect. 5 describes the catalogue itself.  In
the last section we give our conclusions. The catalogue of
superclusters is available electronically at the web-site
\texttt{http://www.aai.ee/$\sim$maret/2dfscl.html}.

\begin{table}[ht]
      \caption[]{Data on comparison samples}
         \label{Tab2}
\begin{center}
         \begin{tabular}{lccrrc}
            \hline\hline
            \noalign{\smallskip}
Sample   & $N_{gal}$ & $r_0$ & $D_0$& $N_{scl}$ & $V$ \\
             & & \Mpc & & & $10^6~(Mpc/h)^3$\\
            \noalign{\smallskip}
            \hline
            \noalign{\smallskip}

Mill.A8   &   8964936 & 8  & 5.0 & 1444 & 125 \\
Mill.A6   &   8964936 & 6  & 6.3 & 1802 & 125 \\
Mill.A4   &   8964936 & 4  & 7.6 & 2244 & 125 \\
Mill.A1   &   8964936 & 0.5& 8.75&32802 & 125 \\
Mill.F8   &   2094187 & 8  & 5.0  & 1734 & 125 \\
Mill.V8   &   1336622 & 8  & 5.0  & 1325 & 125 \\

           \noalign{\smallskip}
            \hline
         \end{tabular}\\
\end{center}
\end{table}

\section{Data}

In this paper we have used the 2dFGRS final release (Colless \etal
\cite{col01}, \cite{col03}) that contains 245591 galaxies. This survey
has allowed the 2dFGRS Team and many others to estimate fundamental
cosmological parameters and to study intrinsic properties of galaxies
in various cosmological environments (see Lahav (\cite{lahav04} and
\cite{lahav05} for recent reviews).  The survey consists of two main
areas in the Northern and Southern galactic hemispheres within the
coordinate limits given in Table~\ref{Tab1}.  The 2dF sample becomes
very diluted at large distances, thus we restrict our sample to a
redshift limit $z=0.2$; we apply a lower limit $z \geq 0.009$ to avoid
the confusion with unclassified objects and stars.  In
Table~\ref{Tab1} $N_{gal}$ is the number of galaxies, $N_{scl}$ is the
number of superclusters found, and $V$ is the volume covered
by the sample (in units $10^6~(Mpc/h)^3$).  These numbers are based
upon version B 
of the catalogue of groups by Tago et al (\cite{tago06}, hereafter
T06). This version of the group catalogue was found using the
Friend-of-Friend (FoF) method with a linking length, which increased
slightly with distance, as suggested by the study of the behaviour of
groups with distance (for details see T06).

The catalogue of groups and single galaxies of T06 gives for all
galaxies equatorial coordinates (for epoch 2000), the ${\rm b_j}$
magnitudes, the morphological parameter $\eta$, the observed absolute
magnitude (and the respective luminosity in Solar units, $L_{obs}$),
and the estimated total luminosity, $L_{tot}$, also in Solar
units. All magnitudes are given in the ${\rm b_j}$ photometric system.

Galaxies were included in the 2dFGRS, if their corrected apparent
magnitude ${\rm b_j}$ lay in the interval from $m_1 = 13.5$ to $m_2 =
19.45$. Actually the faint limit $m_2$ varies  from
field to field.  In calculation of the weights these deviations have
been taken into account, as well as the fraction of observed galaxies
among all galaxies up to the fixed magnitude limit, this fraction is
typically about 0.9, while in rare cases it might become very
small. In such cases, to avoid too high values of respective
corrections, we have applied the completeness correction only 
when the completeness is higher than 0.5, otherwise we assumed a value
1, i.e. no completeness correction was applied.

For comparison we used simulated galaxy samples of the Millennium
Simulation by Springel et al. (\cite{springel05}).  Data on comparison
samples are shown in Table~\ref{Tab2} (for details see Sect. 4).

\begin{figure*}[ht]
\centering
\resizebox{0.48\textwidth}{!}{\includegraphics*{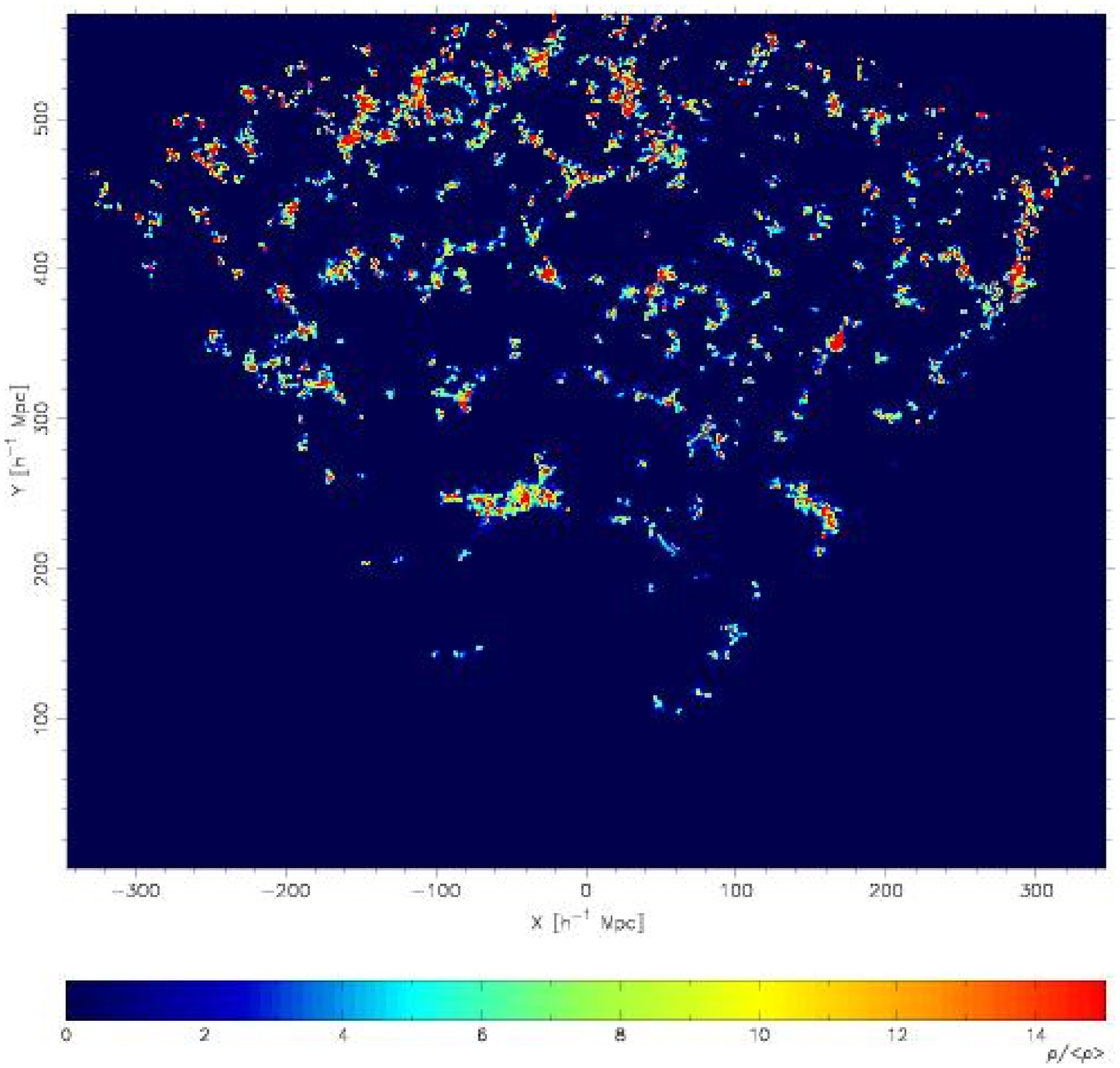}}
\resizebox{0.48\textwidth}{!}{\includegraphics*{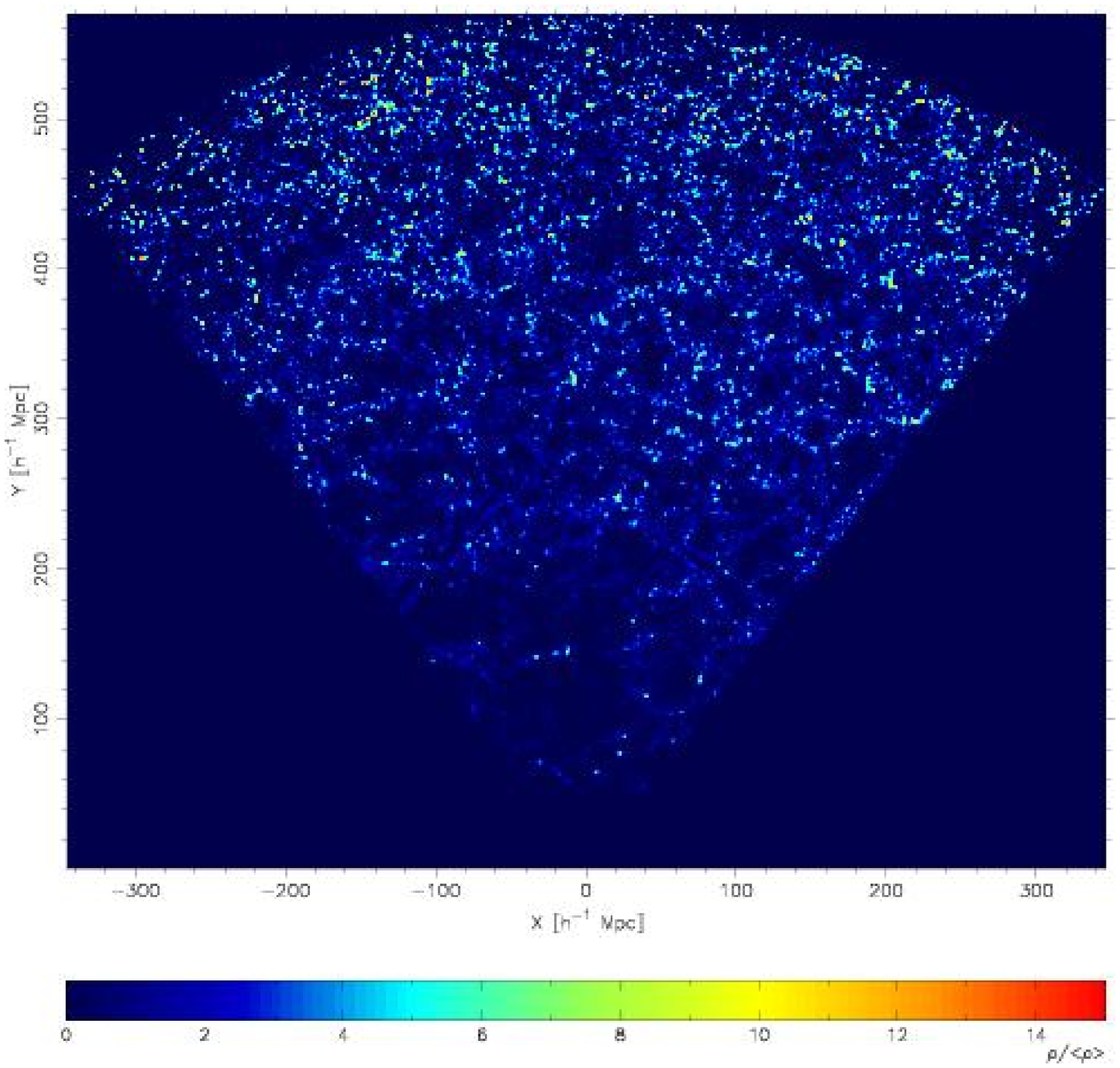}}\\
\resizebox{0.48\textwidth}{!}{\includegraphics*{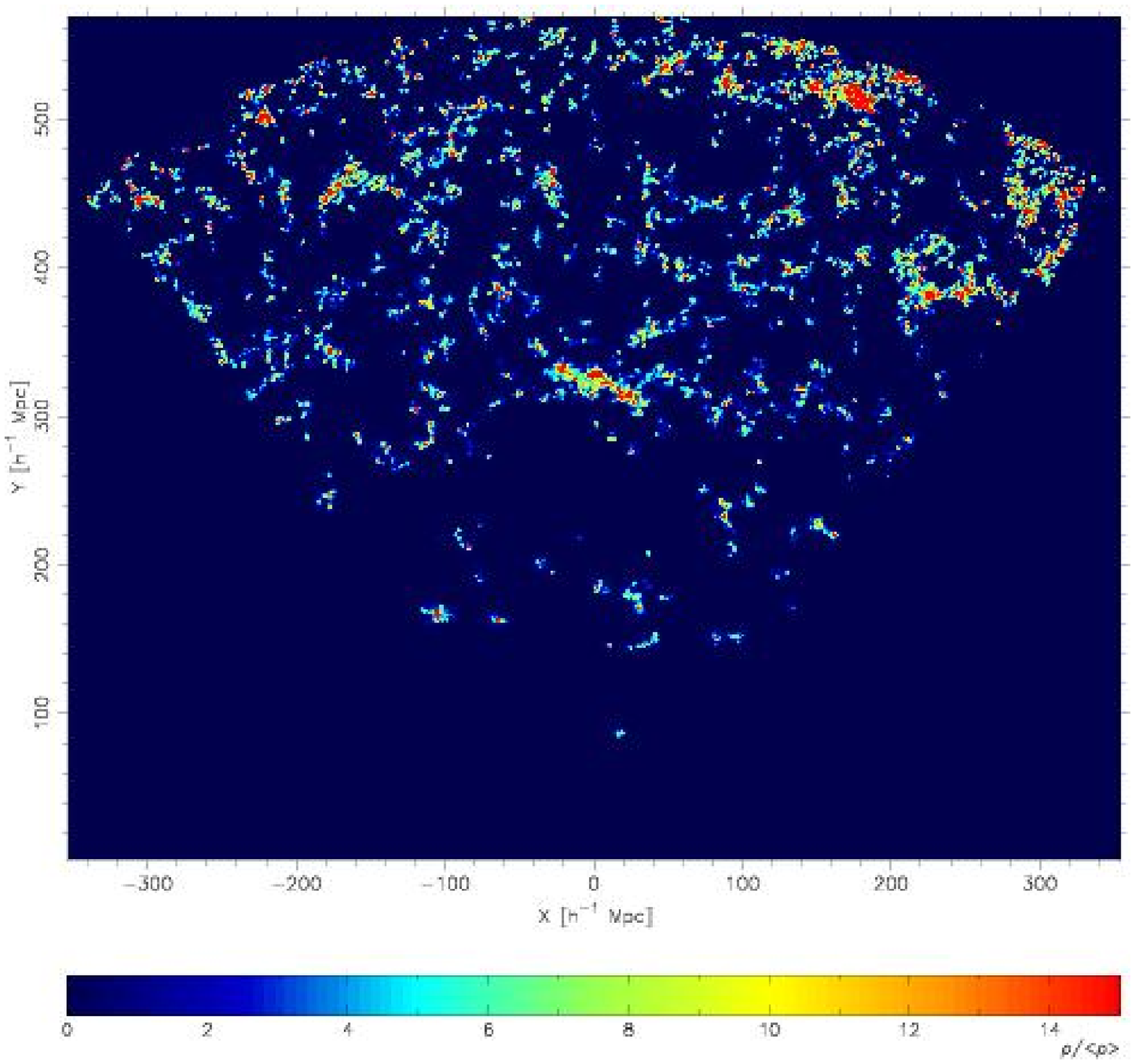}}
\resizebox{0.48\textwidth}{!}{\includegraphics*{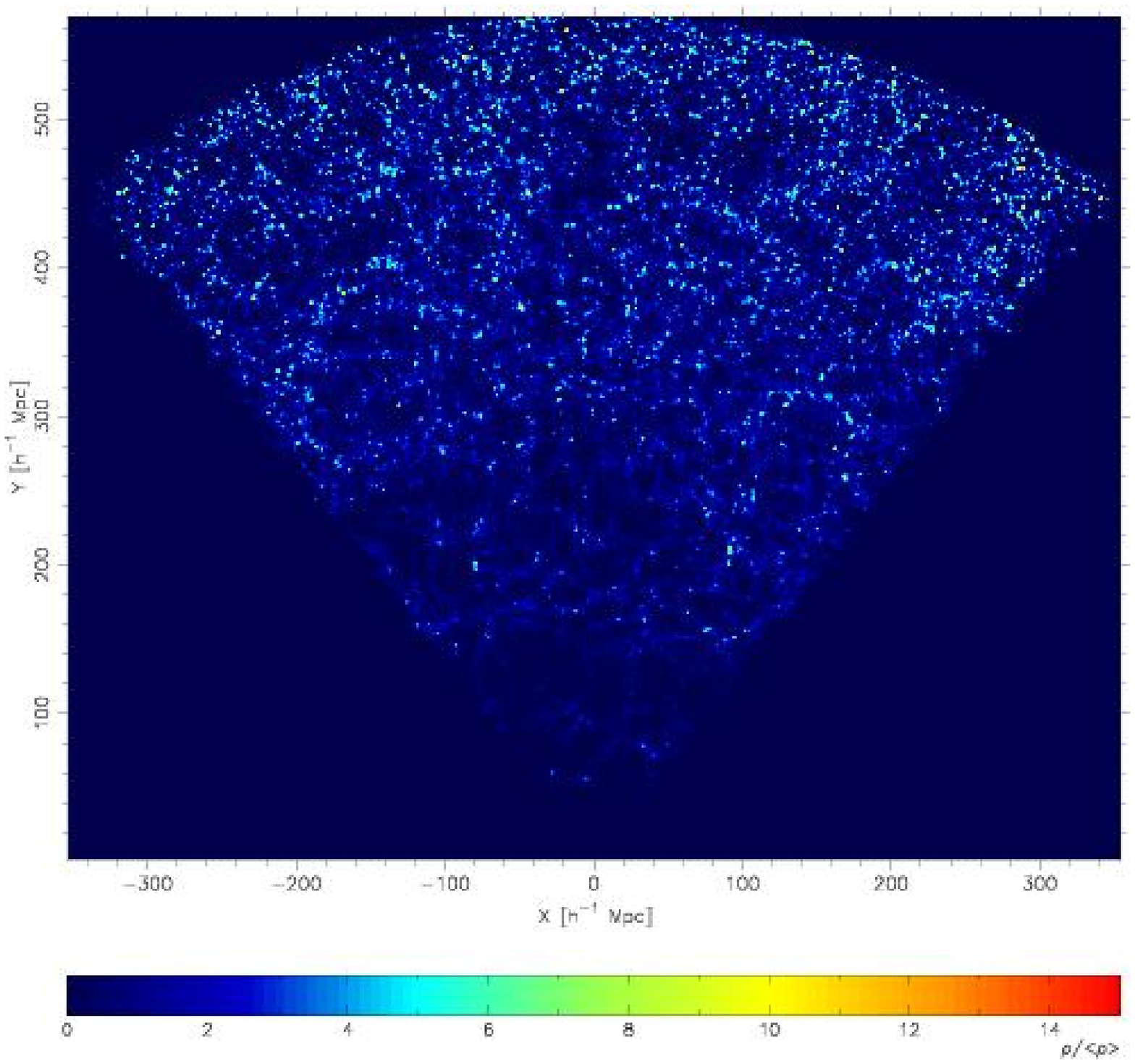}}
\caption{The high-resolution 2-dimensional density field of the
  Northern and Southern parts of the 2dF redshift survey. Upper panels
  show the Northern region, and lower panels the Southern region.  In
  left panels only galaxies and galaxy systems in high-density regions
  are shown, in right panels only galaxies and galaxy systems in
  low-density regions.  The threshold density between high- and
  low-density objects is 4.5 in units of the mean density, smoothed on
  scale 8~\Mpc. The samples are conical, i.e. its thickness increases
  with distance, thus on large distance from the observer we see many 
  more systems of galaxies.  }
\label{fig:1}
\end{figure*}

\section{Superclusters in the cosmic web}

\subsection{Definition of superclusters}

Superclusters have been defined so far either as {\em ``clusters of
clusters''} using catalogues of clusters of galaxies, following Abell
(\cite{abell}, \cite{abell61}), or as high-density regions in the
galaxy distribution, following the pioneering study by de Vaucouleurs
(\cite{deV53}) of the {\em Local Supergalaxy}. Nearby superclusters
have been found mostly on the basis of combined galaxy and cluster
data (J\~oeveer, Einasto, Tago \cite{jet78}, Gregory \& Thompson
\cite{gt78}, Fleenor et al. \cite{fleenor05}, Proust et
al. \cite{proust06a}, \cite{proust06b}, Ragone et
al. \cite{ragone06}).  Until recently, more distant superclusters have
been found almost exclusively on the basis of catalogues of rich
clusters of galaxies by Abell (\cite{abell}) and Abell et
al. (\cite{aco}). Only in recent years distant superclusters have been
found using new deep redshift surveys of galaxies, such as the Las
Campanas Redshift Survey, the 2dF Galaxy Redshift Survey, and the
Sloan Digital Sky Survey (E03a, E03b, Basilakos \cite{bas03}, Erdogdu
et al. \cite{erd04} and Porter \& Raychaudhury \cite{pr05}).

As in previous supercluster searches we are confronted with the problem of 
how to define superclusters. To visualize  the problem  we show
in Fig.~\ref{fig:1} 2-dimensional projections of the 2dF Redshift
Survey Northern and Southern regions. In these plots luminosity density
was found using Gaussian smoothing with rms scale 0.8~\Mpc, 
survey volumes were projected onto a great circle
through respective regions of the sky, and the regions were rotated in
order have  symmetrical areas around the vertical axis.  Galaxies and
galaxy systems located in different global environments are plotted
separately: the left panels show only systems located in high-density
environment, and the right panels show only systems in low-density
environment.  High- and low-density regions are defined by the
low-resolution density field smoothed with Epanechnikov kernel of
radius 8~\Mpc; a threshold density 4.5 was applied in the mean density
units. 

The comparison of left and right panels shows the presence of a
striking contrast between galaxy systems in high- and low-density
regions. Luminous systems in high-density regions are fairly
compact; they have been conventionally classified as superclusters of
galaxies. These systems are well isolated from each other. The
majority of these high-density systems are fairly small in size.  As
we shall see below, these small systems contain only 1 -- 2 clusters
of galaxies and resemble in structure systems like the Local and the
Coma Superclusters.  We see also some very rich superclusters: in the
Northern region the supercluster SCL126 (Einasto et al. \cite{e1997})
or the Sloan Great Wall (Nichol et al. \cite{nic06} and references
therein);  and in the Southern region the Sculptor Supercluster SCL9
(Einasto et al. \cite{e1997}).  

In contrast, galaxy systems in the low-density region form an almost
continuous network of small galaxy filaments.  Faint galaxy systems
are seen even within large low-density regions (cosmic voids). Most
importantly, the distribution of galaxies in space is almost
continuous: faint galaxy bridges join groups and clusters, and thus it
is a matter of convention, where to put the border between
superclusters and poorer galaxy systems.

Traditionally galaxy systems of various scale have been selected from
the cosmic web using quantitative methods, such as the
Friends-of-Friends (FoF) method or the Density Field (DF) method.  In
the first case neighbours of galaxies or clusters are searched using a
fixed or variable search radius. This method is very common in
searching systems of particles in numerical simulations, where all
particles have identical masses. The variant with variable search
radius has been successfully employed in the compilation of catalogues
of groups of galaxies. For the 2dFGRS such catalogues have been
published by Eke et al. (\cite{eke04}) and Tago et al.(T06).  The FoF
method was also used by Berlind et al. (\cite{berlind06}) to find
groups in the SDSS survey, by Einasto et al. (\cite{ e1994},
~\cite{e2001}) in the compilation of the Abell supercluster
catalogues, and by Wray et al. (\cite{wray06}) to find superclusters
in numerical simulations.  This method is simple and straightforward
and especially suitable for volume limited samples, such as the sample
of Abell clusters and similar samples of simulated dark matter haloes.

The FoF method has the disadvantage that objects of different
luminosity (or mass) are treated identically.  Galaxy systems contain
galaxies of very different luminosity from dwarf galaxies to luminous
giant galaxies.  Thus, using the FoF method, it is difficult to make a
clear distinction between poor and rich galaxy systems, if their
number density of galaxies is similar. The second problem of the FoF
method is the complication in using neighbours: the method is simple
if a constant linking length (neighbour search radius) is used, but
the price for this simplicity is the elimination of faint galaxies
from the analysis, in order to get a volume limited galaxy sample.

To overcome these difficulties, the DF method can be used.  Here
luminosities of galaxies are taken into account, both in the search of
galaxy systems, as well as in the determination of their properties.
The second advantage of the DF method is the possibility to make
allowance for completeness and in this way to restore unbiased values
of group (and supercluster) total luminosities.

There exists several variants of the density field method to
investigate properties of the distribution of galaxies.  Basilakos et
al. (\cite{bpr01}) compiled a catalogue of superclusters using the
PSCz flux limited galaxy catalogue, using cell sizes equal to the
smoothing radius, 5~\Mpc\ and 10~\Mpc, for galaxy samples of maximal
distance 150 and 240~\Mpc, respectively.  The use of a fairly large cell
size introduces a bias to the density field, which has been corrected.
Another variant of the density smoothing is the use of the Wiener
Filtering technique, recently applied to the 2dFGRS to identify
superclusters and voids by Erdogdu et al. (\cite{erd04}).  The data are 
covered by a grid whose cells grow in size with increasing distance from 
the observer. Their ``target cell width'' is set to 10~\Mpc\ at the
mean redshift giving a smaller smoothing window for all objects closer
and a much larger for galaxies farther away from us.

Our goal is to find superclusters of galaxies, poor and rich, at all
distances from the observer until a certain limiting distance. To
achieve this goal the selection procedure must be the same for all
distances from the observer.  For this reason we shall use constant
cell size and constant smoothing radius over the whole sample. Of
course, random errors of some quantities increase with distance, but
we want to suppress systematic bias as much as possible.  This allows
the identification of smaller systems at all distances from the
observer.

The key element in our scheme is the restoration of the expected total
luminosity of superclusters as accurately as possible.  This goal can
be achieved using the weight for galaxies in the calculation of the density
field.  We have used this approach in estimating total luminosities of
superclusters of the Las Campanas Survey and Sloan Early Data Release
(E03a, E03b).  We shall describe the estimation of expected total
luminosities in the next section.

To apply the DF method the luminosity density field is calculated using
an appropriate kernel, cell size and smoothing length.  An additional
parameter which influences the sample, is the threshold density to
separate superclusters from poorer galaxy systems. It has the same
meaning as the linking length in the FoF method.  This is the key
parameter which makes a clear distinction between rich and poor galaxy
systems, and its influence is illustrated in Fig.~\ref{fig:1}.  

Additionally a certain minimal radius (or volume) of objects must be
fixed to avoid the inclusion of noise (too small systems) in our
sample. And, finally, certain distance limits must be used for the
whole sample to restrict the study to a region covered by observations
with a sufficient spatial density of objects.  The collection of all
these selection parameters defines the final sample of superclusters.

\subsection{Calculation of expected total luminosities of galaxies}

Due to the selection of galaxies in a fixed apparent magnitude
interval the observational window in absolute magnitudes shifts toward
higher luminosities when the distance of the galaxies increases. This is
the major selection effect in all flux-limited catalogues of
galaxies. Due to this selection effect the number of galaxies seen in
the visibility window decreases.  When calculating estimated total
luminosities of galaxies (and groups) we must take this effect into
account.

We regard every galaxy as a visible member of a group or cluster
within the visible range of absolute magnitudes, $M_1$ and $M_2$,
corresponding to the observational window of apparent magnitudes at
the distance of the galaxy.  To calculate total luminosities of groups
we have to find the estimated total luminosity per one visible galaxy,
taking into account galaxies outside of the visibility window.  This
estimated total luminosity is calculated as follows (E03b)
\begin{equation}
L_{tot} = L_{obs} W_L,
\label{eq:ldens}
\end{equation}
where $L_{obs}=L_{\odot }10^{0.4\times (M_{\odot }-M)}$ is the
luminosity of a visible galaxy of an absolute magnitude $M$, and
\begin{equation}
W_L =  {\frac{\int_0^\infty L \phi
(L)dL}{\int_{L_1}^{L_2} L \phi (L)dL}}
\label{eq:weight}
\end{equation}
is the luminosity-density weight (the ratio of the expected total
luminosity to the expected luminosity in the visibility window).  In
the last equation $L_i=L_{\odot} 10^{0.4\times (M_{\odot }-M_i)}$ are
the luminosity limits of the observational window, corresponding to
the absolute magnitude limits of the window $M_i$, and $M_{\odot }$ is
the absolute magnitude of the Sun.  In the calculation of weights we
assumed that galaxy luminosities are distributed according to the
Schechter (\cite{S76}) luminosity function:
\begin{equation}
\phi (L) dL \propto (L/L^{*})^\alpha \exp {(-L/L^{*})}d(L/L^{*}),
\label{eq:schechter}
\end{equation}
where $\alpha $ and $L^{*}$ are parameters.  Instead of $L^{*}$ the
corresponding absolute magnitude $M^{*} - 5\log_{10} h$ is often used.  We
take $M_{\odot} = 5.33$ in the ${\rm b_j}$ band.  In
calculation of luminosities we used the $k + e$-corrections according to
Norberg \etal (\cite{2002MNRAS.336..907N}).

 \begin{figure}[ht]
\centering
\resizebox{.48\textwidth}{!}{\includegraphics*{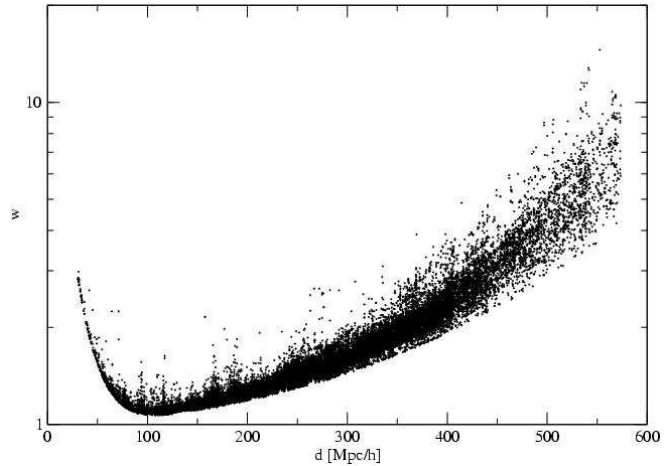}}
\\
\caption{Weights of galaxies to correct observed luminosities for
  calculation of expected  total luminosities
  of superclusters.  
}
\label{fig:02}
\end{figure}

The weights used to calculate estimated total luminosities of
superclusters are shown in the Fig.~\ref{fig:02}.  What is important
here is not only the absence of faint members of groups at large
distance, but also the absence of faint groups.  In this paper we are
interested in the total luminosities of large systems (superclusters),
thus in calculation of estimated total luminosities we use the set of
Schechter parameters $\alpha_1 = -1.21$, $M^{*}_1 - 5\log_{10} h =
-19.66$, as found by Norberg \etal\ for the whole 2dF galaxy sample.
Our calculations show that this set of Schechter parameters yields
total mean luminosity density which is approximately independent of
the distance from the observer, as expected for a fair sample of the
Universe.

\begin{figure*}[ht]
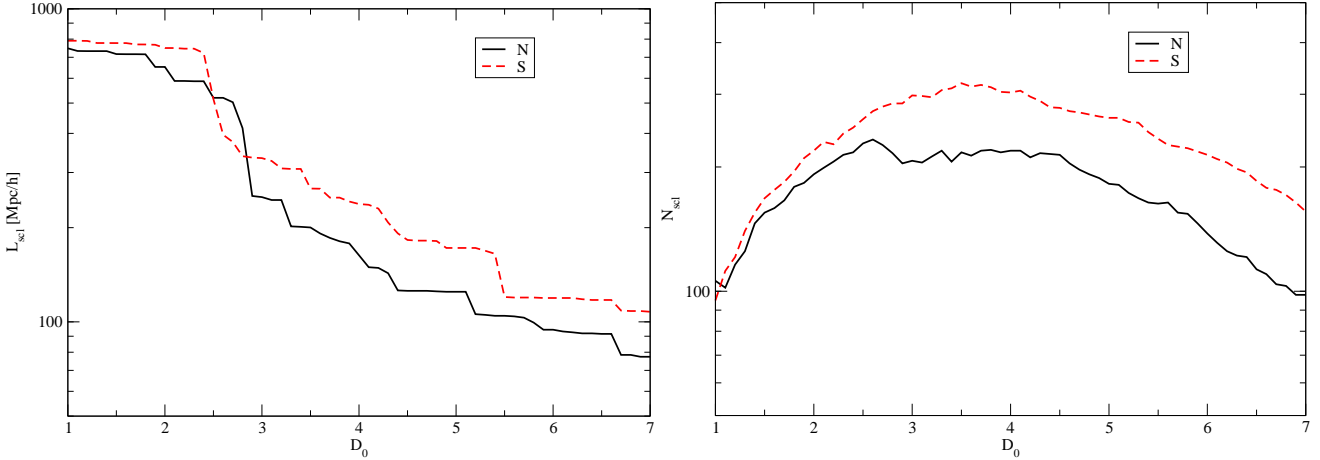

\centering
\resizebox{.48\textwidth}{!}{\includegraphics*{2df_NS_Lscl-Do.eps}}
\resizebox{.48\textwidth}{!}{\includegraphics*{2df_NS_Nscl-Do.eps}}
\\
\caption{Left panel: the maximal diameter of the largest supercluster as a
  function of the threshold density. Right panel: the number of superclusters
  found for various threshold density values.  }
\label{fig:03}
\end{figure*}

\subsection{Preliminary study of 2dFGRS superclusters}

We have compiled our 2dFGRS supercluster sample using three steps:
1) a preliminary study of 2dF superclusters to explore the selection
parameters and to find the most suitable method to select
superclusters; 2) investigation of superclusters in simulated galaxy
samples for further analysis of selection parameters and possible
biases and errors; 3) selection of the final 2dFGRS supercluster
catalogue using parameters chosen during the preliminary study.  In
the preliminary phase we applied both the FoF and DF methods.

As in the compilation of the group catalogue by T06 we accept the
upper limit of redshift of galaxies used in the supercluster search
$z=0.2$, corresponding to a distance of $d=575$~\Mpc.  In the calculating 
distances we use a flat cosmological model with the parameters: matter
density $\Omega_m = 0.27$, dark energy density $\Omega_{\Lambda} =
0.73$ (both in units of the critical cosmological density), and the
mass variance on 8~\Mpc\ scale in linear theory $\sigma_8 = 0.84$.
Here and elsewhere $h$ is the present-day dimensionless Hubble
constant in units of 100 km s$^{-1}$ Mpc$^{-1}$.

The FoF method is simple when absolute magnitude (volume) limited
galaxy samples are used.  In this case one can use a constant linking
length over the whole sample to find superclusters.  We tried two
limiting absolute magnitudes, $-19.0$ and $-19.5$, with distance
limits 400 and 520~\Mpc, respectively. A lower limit of the number of
galaxies in superclusters of 100 was chosen.  Superclusters were selected
in the Northern and Southern regions; selection limits in coordinates
are given in Table~\ref{Tab1}.

For the DF method we used a cell size of 1~\Mpc.  This is the
characteristic size of compact galaxy systems -- groups and clusters.
Using this cell size and a small smoothing length 0.8~\Mpc\ it was
possible to follow the distribution of compact galaxy systems
(clusters) (E03a, E03b).  To characterize the global environment of
galaxies a smoothing with characteristic scale 8 -- 10 \Mpc\ has been
applied, using either a Gaussian or an Epanechnikov kernel, see De
Propis et al. (\cite{dep03}), Croton et al. (\cite{cr04}) and Einasto
et al.  (\cite{ein05b}, hereafter E05b).  To avoid excessive smoothing
with large wings we used the Gaussian smoothing only to calculate the
high-resolution density field with rms scale 0.8~\Mpc.  To find the
low-resolution field we used the Epanechnikov kernel
\begin{equation}
k(r) ={3 \over {4r_0}} (1 - (r/r_0)^2),
\end{equation}
where $r_0$ is the limiting radius for smoothing.  We accepted in the
following analysis the  radius 8~\Mpc.

The next step in the selection of superclusters is the proper choice
of the threshold density $D_0$ to separate high and low-density galaxy
systems.  Following E03a we compiled supercluster catalogues in a wide
range of threshold densities from 1 to 7 in units of the mean luminosity
density of the sample.  For each threshold density value we found the
number of superclusters, $N_{scl}$, and calculated the maximal
diameter of the largest system found, $L_{scl}$ (see Fig.~\ref{fig:03}).
Detailed supercluster lists and their properties were calculated for
several threshold densities in the range 4$\dots$5 (in units of the
mean density).

\begin{figure*}[ht]
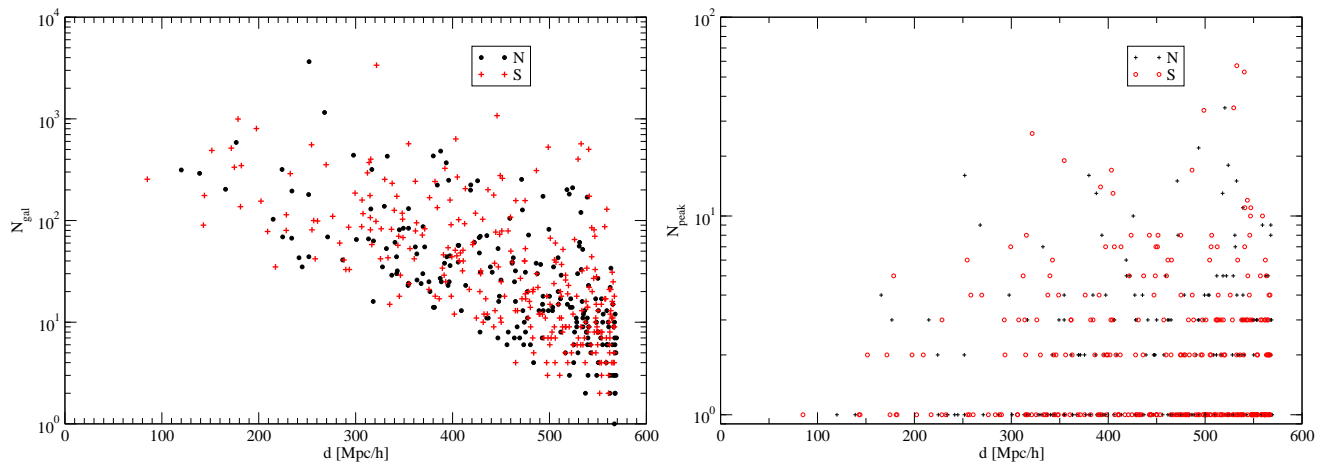

\centering
\resizebox{.48\textwidth}{!}{\includegraphics*{2df_NS8_Ngal-dist.eps}}
\resizebox{.48\textwidth}{!}{\includegraphics*{2df_NS8_Npeak-dist.eps}}
\\
\caption{Left panel: the number of galaxies in superclusters at
  various distance from the observer. Right panel: the multiplicity
  of superclusters (defined as the number of DF-clusters) as a
  function of distance from the observer.  }
\label{fig:04}
\end{figure*}

Finally we have to fix the minimal volume (or radius) of systems to be
considered as superclusters.  This choice is important  in order to have a
difference between compact galaxy systems, such as groups and
clusters, and more extended systems -- i.e., superclusters.  We take into
account the fact that all compact systems transform to extended
objects after smoothing. In our previous analysis (E03a and E03b) we
used Gaussian smoothing with rms scale 10 \Mpc, and the limiting
radius of the smallest system to be considered as a supercluster 5.04~
\Mpc; this radius corresponds to a system of volume 535~(\Mpc)$^3$.
When using an Epanechnikov kernel with radius $\approx 8$~\Mpc\ we can
use a smaller limiting radius.  Taking these considerations into
account we used in our preliminary study 3.63~\Mpc\ as the limiting
radius, which corresponds to a limiting volume of 200~(\Mpc)$^3$.

\subsection{DF-clusters in superclusters}

To get an idea of selection effects we show in Fig.~\ref{fig:04} the
number of galaxies in the superclusters.  This number was found by
searching for galaxies which lie in the volume above the threshold
density level.  As we see, the number decreases exponentially with
distance.  This effect is expected, since galaxies in the 2dFGRS
sample are flux-limited, and at larger distances faint galaxies fall
outside the observational window. A similar dependence is observed
for groups of galaxies of the T06 sample for the same reason:
faint distant groups cannot be detected.

This example shows that we cannot use the number of galaxies or groups
as the richness criterion of superclusters.  Instead of galaxies or
groups we can use DF-clusters to characterize the richness of
superclusters.  The density field is corrected for selection effects
using appropriate weights. This approach has been used by E03a and
E03b, where lists of DF superclusters and DF-clusters have been
compiled. We have followed this experience and have found lists of
DF-clusters for all our samples. 

To find DF-clusters we used the low-resolution density field, since it
averages over the cluster environment, thus giving higher weight to
clusters which are located in a high-density environment. In practical
terms, all density peaks of the low-resolution density field were
located, having a peak density a bit higher than the threshold density
used in the supercluster search. We used minimal peak density 5.0 in
units of the mean density. The DF-cluster is characterised by its peak
density and its integrated peak density found by summing luminosity
densities of all cells around the peak together with the central cell
in 27 cells.  Additionally we searched for galaxies and groups around
the central peak within relative distance limits $\pm 8$~\Mpc\ from
the central peak.  The smoothed density field integrates luminosities
of galaxies inside the whole sphere of radius equal to the smoothing
radius, $4\pi r_0^3/3 = 2145$~(\Mpc)$^3$, and DF-clusters characterize
the luminosity of the central cluster as well as that of surrounding
galaxies and groups. This sphere contains in nearby regions
200$\dots$1000 galaxies and in most distant regions 3$\dots$50
galaxies.

The analysis shows that poor superclusters contain 1 -- 2 DF-clusters,
i.e. they a similar to the Local and Coma superclusters.  In rich
superclusters the number of DF-clusters is much higher.  The
distribution of the multiplicity of superclusters is shown in the
right panel of Fig.~\ref{fig:04}. We see that the distribution is
practically independent of the distance.  At small distances the
number of high-multiplicity superclusters is smaller, but this is a
volume effect. What is more important, low-multiplicity superclusters
are detected at all distances.

The analysis of properties of superclusters found with the DF and FoF
methods demonstrates that the main properties of rich superclusters
(positions, diameters, total luminosities etc) are rather stable and
do not depend too much on the method to select them.  In most cases it
was possible to make a one-to-one identification of superclusters
found with different methods or sets of selection parameters.  Of
course, in some cases a rich supercluster found with one method was
split into two or more subclusters when a different set of parameters
or method was used.  As the real cosmic web is continuous, such
differences are expected. It is encouraging that these differences
were rather small.  The only major disadvantage of the FoF method is
that a large fraction of the data is not used, since all galaxies
fainter than the magnitude limit are ignored.  Thus the following
analysis was carried out with the DF method only.

\section{Analysis of simulated superclusters}

The final step in our preliminary study is the analysis of
simulated superclusters using the Millennium Simulation.  The use of
simulated superclusters has the advantage that true properties of
model superclusters are known, and the comparison of properties of
superclusters based on full data and simulated 2dF data allows us to
estimate possible errors and biases of real superclusters.

\subsection{Selection effects and biases of the catalogues} 

The major issue in using flux-limited galaxy samples as the 2dFGRS is
the magnitude selection effect. Due to a fixed observational window in
apparent magnitudes the range of absolute magnitudes of galaxies (and
groups selected from the galaxy sample) changes with the distance.
At the far side of the observational sample only very bright galaxies
fall into the visibility window of the sample. Thus the number-density
of galaxies drops with increasing distance dramatically.  This makes
it difficult to estimate the true number of galaxies and groups in
superclusters.

\begin{figure*}[ht]
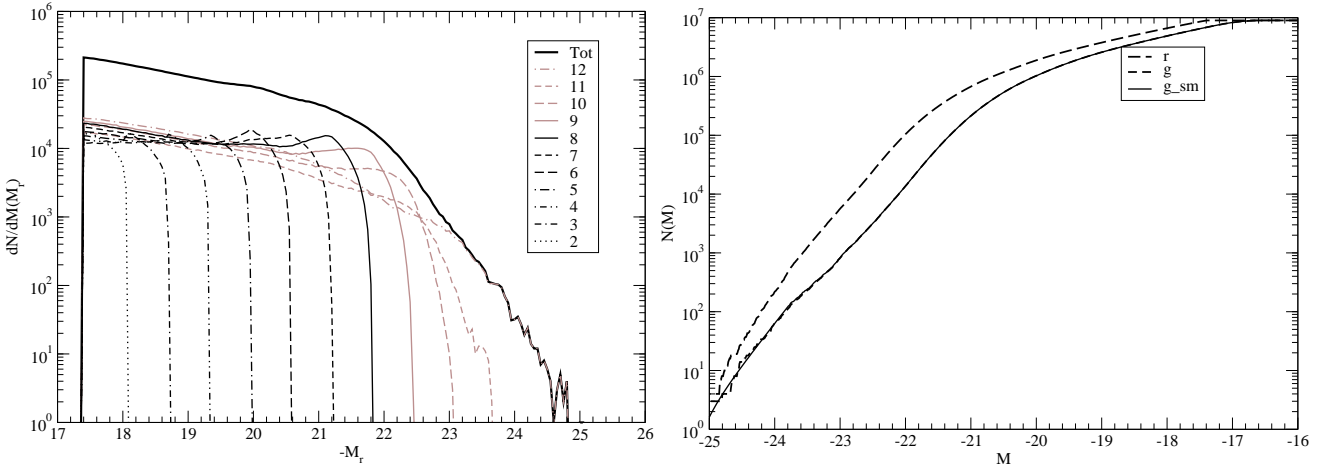

\centering
\resizebox{.48\textwidth}{!}{\includegraphics*{mill_DLF.r-dens.eps}}
\resizebox{.48\textwidth}{!}{\includegraphics*{mill_IntLF-gr.eps}}
\\
\caption{Left panel shows the differential luminosity function (the
  number of galaxies in absolute magnitude bins of $\Delta r =
  0.05$). Thin lines show the luminosity function in different local
  density environment, found with Gaussian smoothing of scale
  0.8~\Mpc; lines labeled 2, 3 $\dots$ correspond to local densities
  in intervals of density logarithm $\log D = -0.75 \dots -0.50, -0.50
  \dots -0.25~\dots$. Bold line shows the whole differential
  luminosity function in the r band.  The right panel shows
  integrated luminosity functions in the {\tt r} and {\tt g} 
  bands. For high luminosities the smoothed approximation of the
  function is plotted.
}
\label{fig:5}
\end{figure*}

To investigate selection effects in compiling the catalogue of groups
of the 2dFGRS Tago et al (T06) used a simple method: nearby real
groups were shifted to larger distance, and the change of the number
of group members was investigated.  In the present paper we shall use
for the study of selection effects simulated galaxies and galaxy
systems found in the Millennium Simulation of the structure
evolution. For details of the model see Springel et
al. (\cite{springel05}), Croton et al. (\cite{cr06}) and Gao et
al. (\cite{gao05}). This simulation was made using modern values of
cosmological parameters in a box of side-length 500~\Mpc, using a very
fine grid (about $2000^3$), and the largest so far number of Dark
Matter particles.  Using semi-analytic methods simulated galaxies were
calculated. The simulated galaxy catalogue contains almost 9 million
objects, for which  positions and velocities are given, as well
as absolute magnitudes in the Sloan Photometric system
({\tt u,g,r,i,z}). The limiting absolute magnitude of the catalogue is
$-17.4$ in the {\tt r} band.

In order to study the influence of the smoothing length we applied
an Epanechnikov kernel with radius 4, 6, and 8~\Mpc\ to find the luminosity
density field; respective models are marked in Table~\ref{Tab2} as
Mill.A4, Mill.A6 and Mill.A8 (A for all galaxies used in calculation
of the density field). Further we simulated the influence of the
observational selection.  We put the observer at the lower left corner
of the sample at coordinates $x = y = z = 0$, calculated distances of
every galaxy from the observer, found apparent magnitudes using
$k-$corrections, and selected galaxies in the observational window of
the 2dFGRS $m_1 = 14.5$, $m_2 = 19.35$; this subsample is designated
as Mill.F8.  To simulate volume-limited galaxy samples we applied a
further limit, $-19.5$, in absolute magnitudes in photometric system g
(close to system ${\rm b_j}$ used in the 2dF Survey); this subsample
is designated Mill.V8 (in the last samples a smoothing radius 8~\Mpc\ was
applied).

\begin{figure*}[ht]
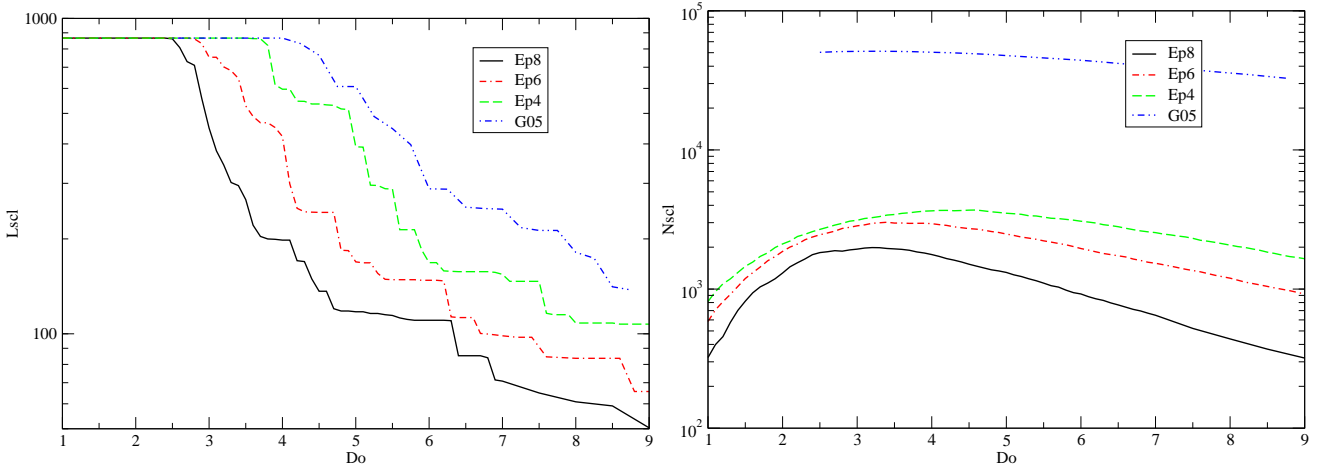

\centering
\resizebox{.48\textwidth}{!}{\includegraphics*{scl_M500ep-all_Lscl-Do.eps}}
\resizebox{.48\textwidth}{!}{\includegraphics*{scl_M500ep-all_Nscl-Do.eps}}
\\
\caption{The length (maximal diameter) and the number of superclusters
  are shown in the left and right panels, respectively, as a function
  of the threshold density $D_0$.  Different lines show data using
  smoothing with an Epanechnikov kernel of radius 8, 6 and 4~\Mpc, and
  Gaussian kernel of scale 0.5~\Mpc.
}
\label{fig:6}
\end{figure*}

\begin{figure*}[ht]
\centering
\resizebox{.48\textwidth}{!}{\includegraphics*{scl_M5ep-all_MeanLscl-Do.eps}}
\resizebox{.48\textwidth}{!}{\includegraphics*{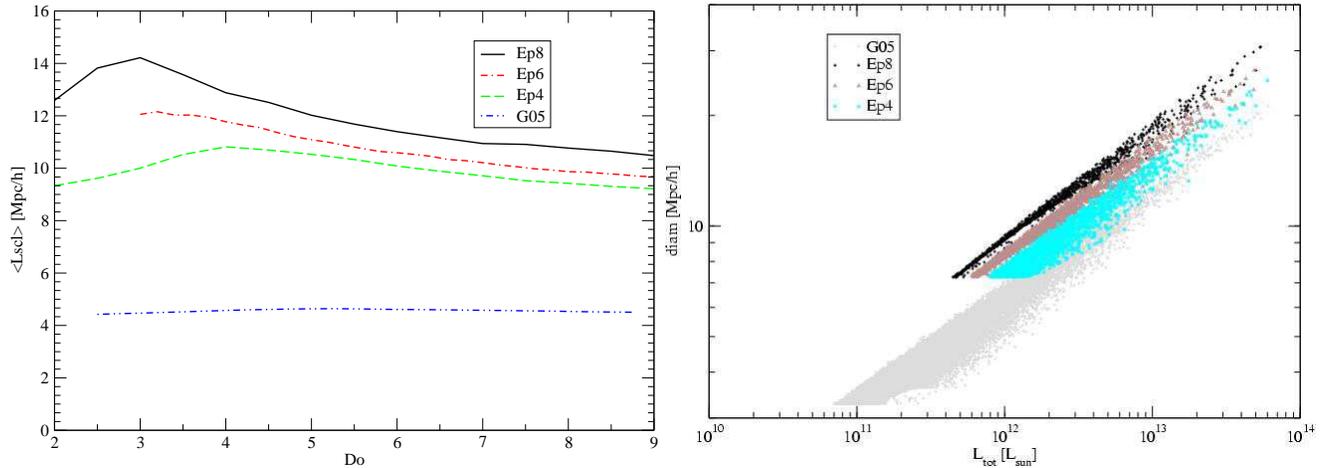}}
\\
\caption{Left panel: the mean length (diameter) of superclusters as a
  function of the threshold density $D_0$.  Right panel: effective
  diameters of superclusters of various total luminosity. Different
  lines and symbols are for samples as in Fig.~\ref{fig:6}.  }
\label{fig:7}
\end{figure*}

One of the first questions to be clarified is: is the
luminosity-density relation, observed in the real Universe, also
incorporated in simulations? Our experience has shown that it is not
always taken into account in simulating galaxies in numerical models.
We calculated the density field with a Gaussian kernel of rms scale
0.5~\Mpc; this variant is designated Mill.A1. Further we found for
every galaxy the local density value at the position of the galaxy,
and calculated the number of galaxies in various absolute magnitude
intervals separately for different local density environment.  In
magnitudes we used a step $\Delta M = 0.05$, and for density we used
constant intervals in the logarithm of the density with step $\Delta
\log D = 0.25$, starting from density value 0.1 in units of the mean
luminosity density.

Results of this study are shown in Fig.~\ref{fig:5}.  There are almost
no galaxies in the first density bin ($\log D = -1.00 \dots
-0.75$). Starting from the second bin each subsequent density bin
contains more brighter galaxies, the number of faint galaxies in each
bin is approximately constant, and the increase of the maximal
luminosity is practically constant, when we move from lower density
bins to higher ones.  In other words, the luminosity-density relation
is built in to the galaxy sample, and we can use the sample to study
supercluster properties.  The integrated luminosity function for the g and
r bands is shown in the right panel of Fig.~\ref{fig:5}.
Due to very large number of galaxies in the sample, the functions are
very smooth, and only for the bright end was it necessary to apply a
linear interpolation of the function (in $\log N -- M$
representation), also shown in Fig.~\ref{fig:5}.  This luminosity
function was used instead of the Schechter law in calculating weights
for galaxies according to Eq.\ref{eq:weight}.

\subsection{The test for variable smoothing length}

We used the density fields calculated with an Epanechnikov kernel with
radius 4, 6, and 8~\Mpc\ to select superclusters in a wide range of
threshold densities from 1 to 9 in units of the mean luminosity
density. For comparison we applied a similar system search also for the
high-resolution density field found with a Gaussian kernel and rms scale
0.5~\Mpc.  In the latter case we used a minimal volume of systems
20~$(Mpc/h)^3$, since this smoothing scale is suitable for the search
of compact galaxy systems, such as groups and clusters.

The length (maximal diameter)  and the number of systems found are
shown in Fig.~\ref{fig:6} for all four subsamples.  As in the case of
real galaxy samples at low threshold density the largest system spans
the whole region. To avoid the inclusion of very large percolating
systems within our model supercluster catalogue, the threshold density has
to be chosen so that the size of the largest system (diameter of the
box around the system along coordinate axes) does not exceed a certain
value of $100 \dots 150$~\Mpc.  We have chosen values given in
Table~\ref{Tab2}, which correspond to the diameter of the largest
supercluster ($\approx 120$~\Mpc). If one wants to get a higher number
of superclusters, then a lower threshold density is to be used, but in
this case the size of the largest system exceeds 200~\Mpc.

In addition to the diameter of the box around the supercluster we have
found also the diameter of the sphere equal to the volume of the
superclusters, by counting cells of size 1 \Mpc\
inside the contour surrounded by threshold density level. We call this
the effective diameter.  Mean values of the effective diameters of
superclusters of all samples are shown in Fig.~\ref{fig:7} for various
threshold density levels.  We see that, in spite of the presence of
very large percolating superclusters at low threshold density levels,
the mean diameters are surprisingly constant.  For our accepted
threshold levels they lie between $10 \dots 12$~\Mpc, for
superclusters of samples Mill.A4 $\dots$ Mill.A8. The mean diameter of
galaxy systems of the sample Mill.A1 is much lower since a lower
limiting volume was used in the compilation of this sample.

The right panel of Fig.~\ref{fig:7} shows the effective diameters of
individual superclusters as a function of their total luminosity
(found by adding luminosity density values inside the threshold
density contour multiplied by the mean luminosity per cell of the
whole sample).  We see that a very close relationship  exists 
between the diameter and the luminosity of the supercluster.  This
close relationship is due to the fact that mean densities of
superclusters vary in rather narrow limits. The strips of points for
various subsamples are shifted with respect to each other: for a given
luminosity the supercluster diameter is larger for larger smoothing
kernels.

\begin{figure}[ht]
\centering
\resizebox{.48\textwidth}{!}{\includegraphics*{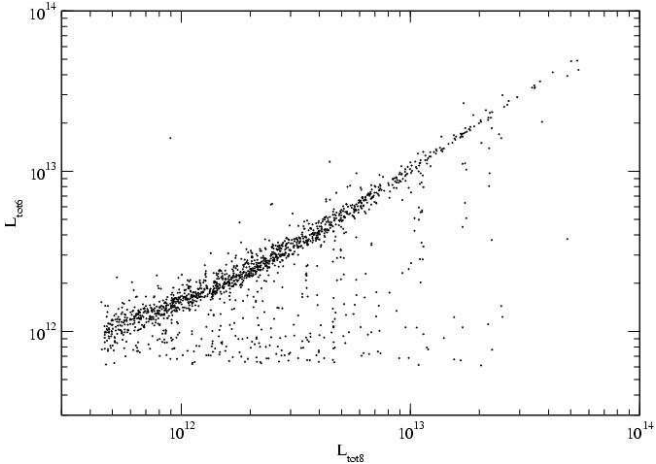}}
\\
\caption{The comparison of total luminosities of superclusters found for
  models Mill.A8 and Mill.A6. 
}
\label{fig:8}
\end{figure}

We have cross-correlated individual superclusters of subsamples
Mill.A8, Mill.A6 and Mill.A4.  For this purpose we find for every
supercluster of subsamples Mill.A6 and Mill.A4 the closest
supercluster of the sample Mill.A8.  In most cases the mutual distance
between such supercluster pairs from different subsamples is close to
zero, i.e. we have found identical objects in both subsamples.  Most
very rich superclusters have almost identical counterparts of close
total luminosity in different subsamples, as seen in Fig.~\ref{fig:8}
where total luminosities of cross-identified superclusters are
compared.  However, the lower the total luminosity of the
supercluster the more often a supercluster in subsample Mill.A8 is
split into two or more units in subsamples Mill.A6 and Mill.A4.  In
these cases luminosities of corresponding superclusters of subsamples
Mill.A6 and Mill.A4 are  lower than in the sample Mill.A8.  This
explains the presence of numerous dots below the main ridge in
Fig.~\ref{fig:8}. 

When we compare the distribution of supercluster luminosities of
various subsamples,  we see that the smaller the smoothing length
in calculation of the density field, the higher the number of
low-mass superclusters in the subsample.  This tendency is seen also
in Fig.~\ref{fig:8}.  It is well-known that bridges between
high-density knots in the galaxy distribution consist of faint
galaxies (due to the density-luminosity relation).  If the smoothing
length is small then these bridges fall below the density threshold
and a galaxy system is considered as consisting of two separate
systems. In other words, the density field becomes  noisier. 

When one uses flux-limited galaxy samples, then at larger distance
from the observer fainter galaxies are not visible and bridges between
high-density knots cannot be detected.  In volume limited samples
fainter galaxies are excluded at all distances from the observer. Thus
in real galaxy samples faint galaxy bridges disappear at large
distance (or everywhere for volume-limited samples).  To avoid a too
noisy density field it is reasonable to use larger smoothing
length. In the following we shall use only supercluster samples found
with 8~\Mpc\ smoothing.

\subsection{Determination of supercluster centres}

It is well-known that rich superclusters are great attractors.  This
effect is very well seen in numerical models, where it is easy to
calculate the potential field. It is natural to identify the centers
of superclusters with centres of deepest potential wells inside the
supercluster. Rich superclusters have several concentration centres
(DF-clusters); the depth of the respective potential wells is
different, and only one has the deepest level. In such cases it is
relatively easy to identify the dynamical center of the supercluster.
The center identification is not so easy in real observational
samples.  To calculate the potential field the respective density
field must be given in a rather large volume. This is easy in
numerical models, but difficult in the real Universe, since even the
largest modern redshift surveys cover relatively thin slices.

\begin{figure}[ht]
\centering
\resizebox{.48\textwidth}{!}{\includegraphics*{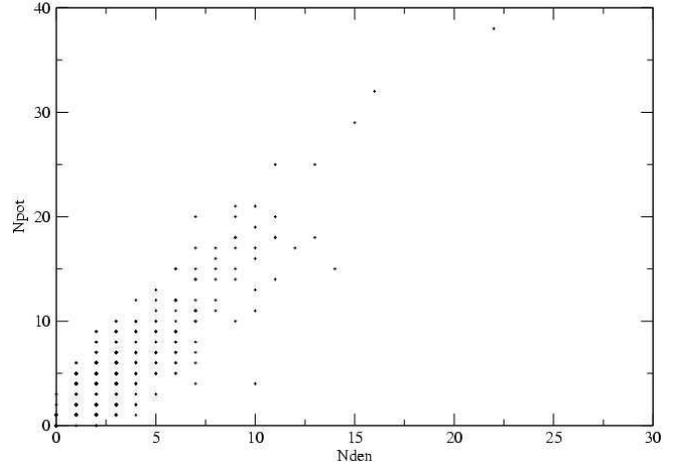}}
\\
\caption{The number of peaks of the density and potential
  field for superclusters of the sample Mill.A8.  
}
\label{fig:9}
\end{figure}

However, it is very easy to identify in real galaxy samples
high-density knots of the density field -- DF-clusters. The problem
is: how well do the positions of high-density knots correlate with positions
of depressions in the potential field?  To study this problem, we
calculated the potential field of the Millennium Survey by
Fourier-transforming the high-resolution density field (model
Mill.A1). Both fields were calculated on a $512^3$ grid.  To have an
impression of the fields they were transformed to FITS format and were
scrutinized using the ds9 viewer (Smithsonian Astrophysical
Observatory Astronomical Data Visualization Application, available for
all major operating systems).  This viewing impression, as well as the
comparison of lists of density peaks and potential field depressions
shows that practically all high-density knots in the density field can
be recognized as depressions in the potential field.

To obtain a more quantitative relationship between maxima (and minima in
case of the potential) of these fields we compared catalogues of
maxima of the high-resolution density field and minima of the
potential field. Also, in our test catalogues of superclusters extrema
of both fields were marked.  This comparison shows that the number of
peaks of both fields in superclusters is close (see Fig.~\ref{fig:9}),
and that in the majority of cases there exists a one-to-one
correspondence between peaks of both fields. In the majority of cases
the highest density peak corresponds to the deepest potential well.
However, in about 10\% superclusters the deepest potential well
coincides not with the highest density peak, but with one the
following peaks.  This occurs mostly in cases where the surrounding
potential field has a considerable slope (even within the
supercluster), so that absolute values of the depth of the potential
well do not always represent the strength of the density peak.  Our
impression is that in these cases the highest density peak suits even
better as the center of the supercluster.

\begin{figure*}[ht]
\centering
\resizebox{.48\textwidth}{!}{\includegraphics*{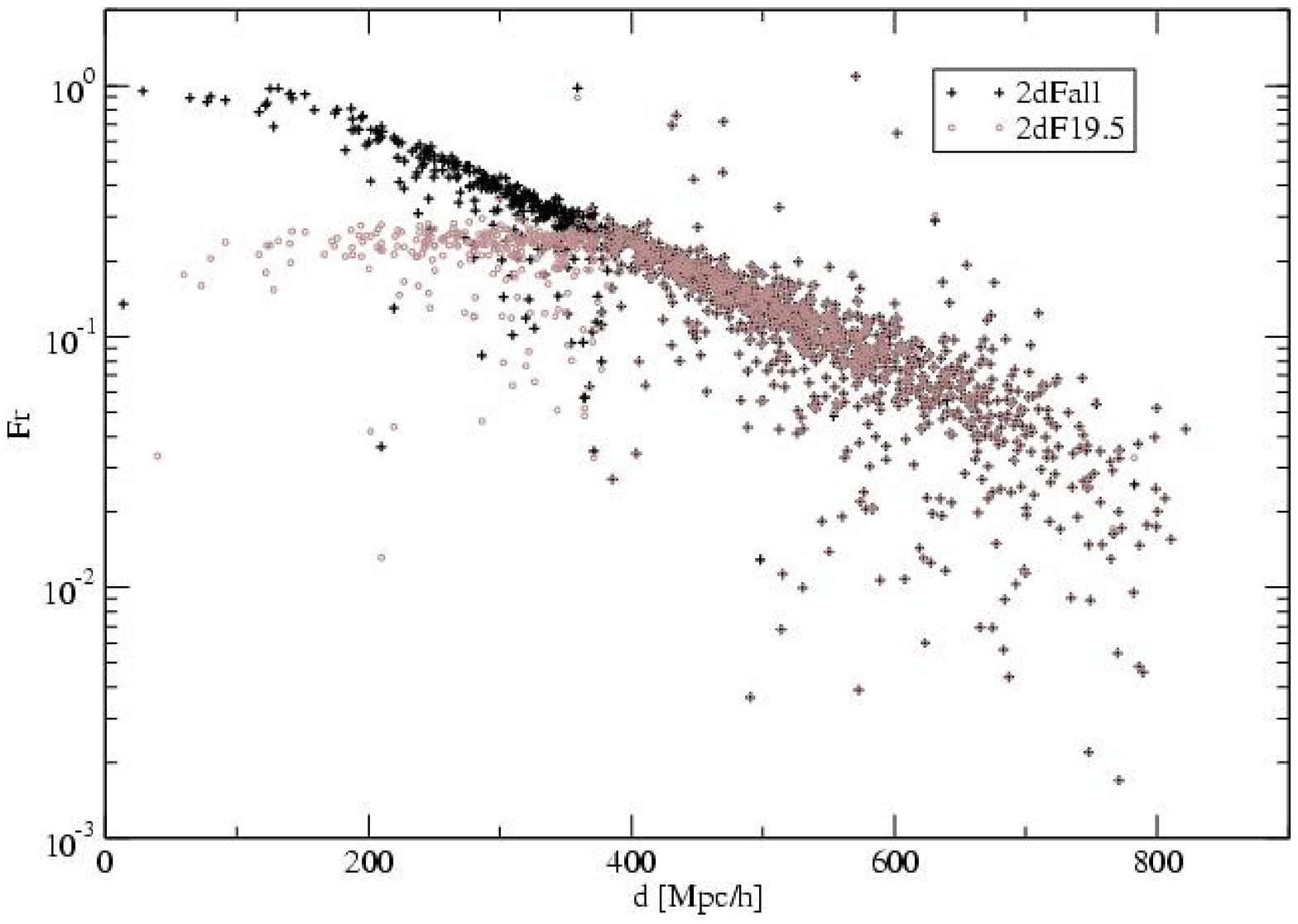}}
\resizebox{.48\textwidth}{!}{\includegraphics*{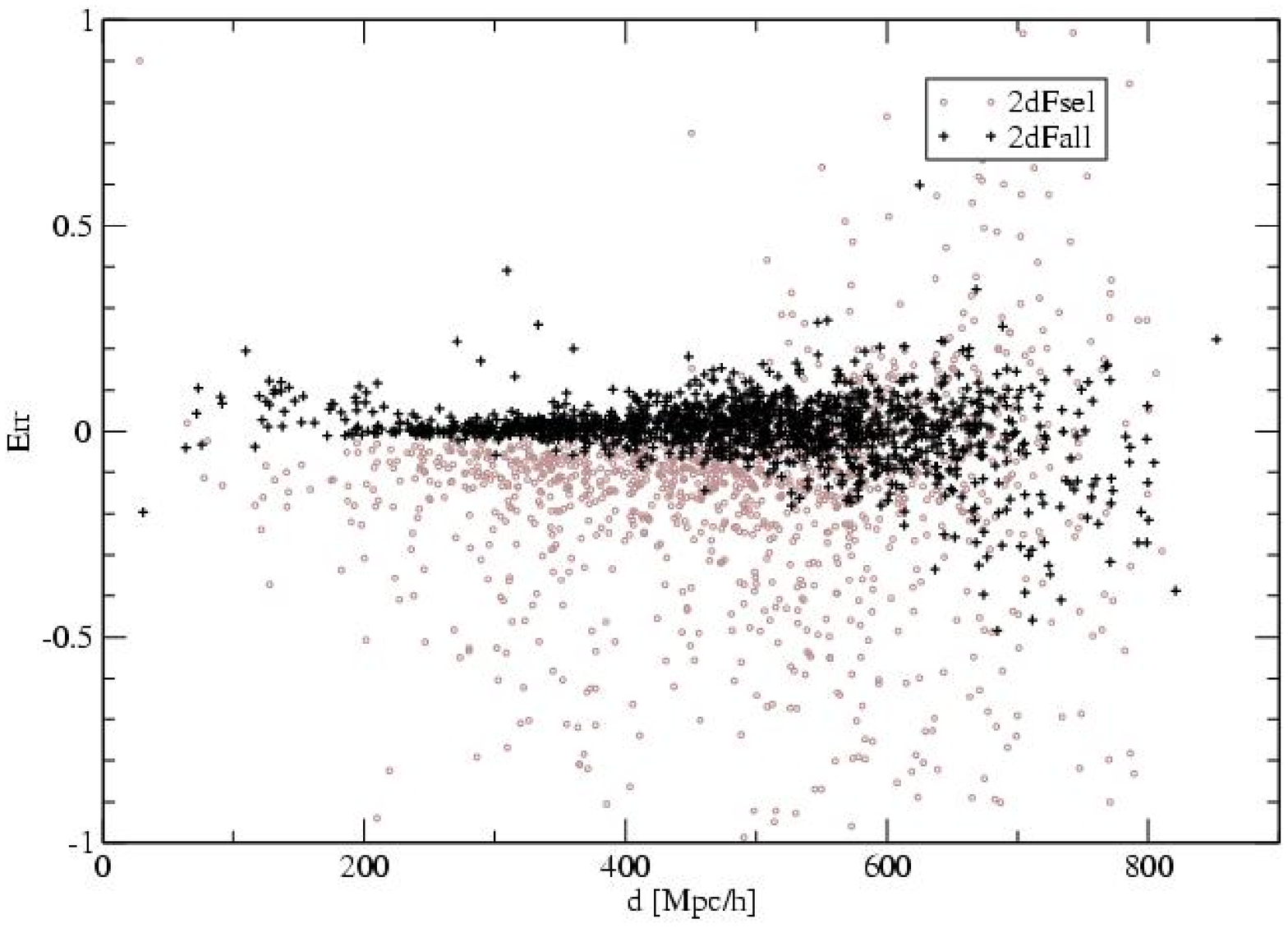}}
\\
\caption{The left panel shows the fraction of superclusters of 2dF
  full and volume limited samples, Mill.F8 and Mill.V8, as a fraction
  of the total sample Mill.A8, for various distance from the
  observer. The right panel shows relative errors
  of total luminosities of superclusters of the sample Mill.F8 
   with respect to the total sample Mill.A8.
  }
\label{fig:10}
\end{figure*}

\begin{figure*}[ht]
\centering
\resizebox{.48\textwidth}{!}{\includegraphics*{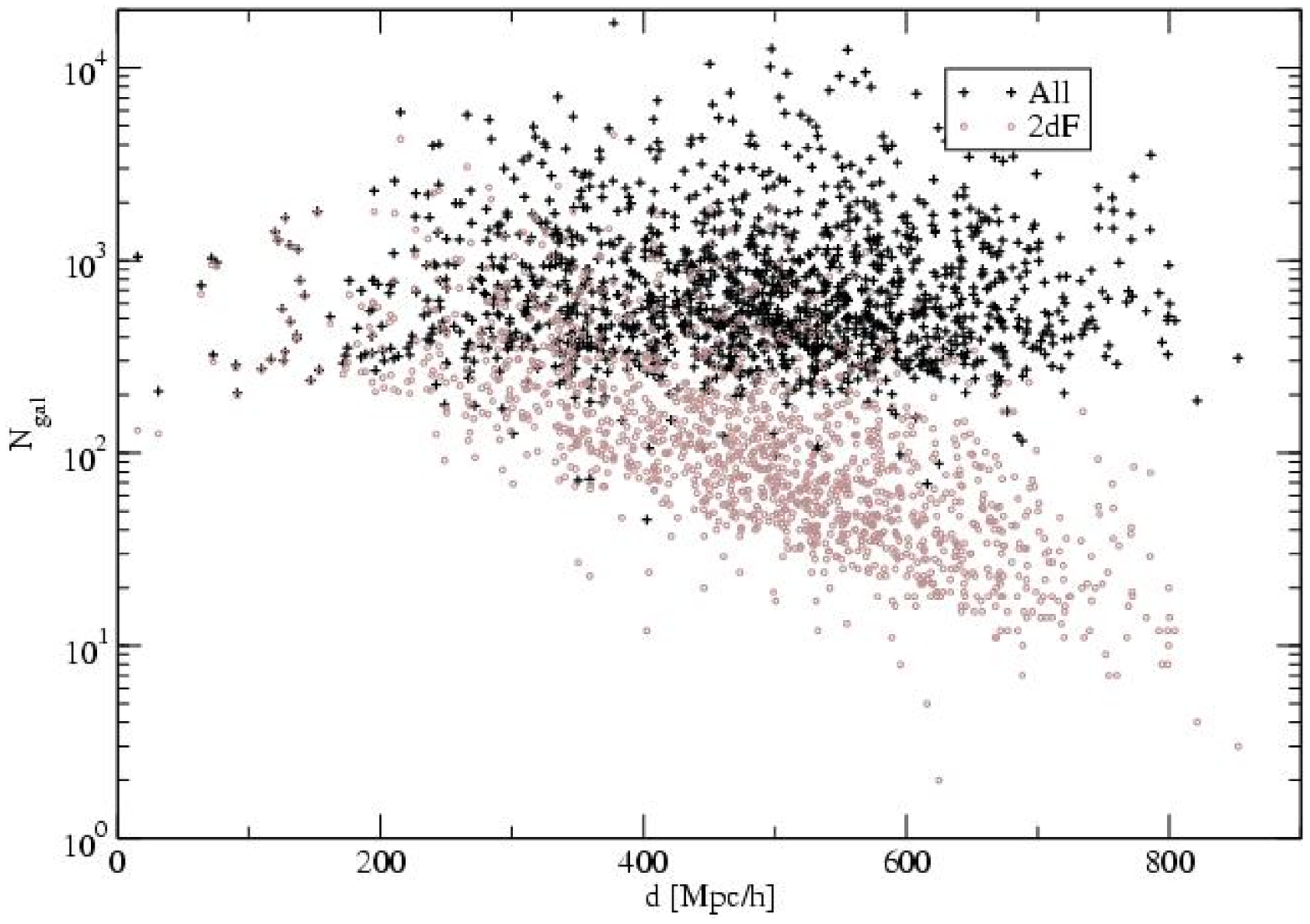}}
\resizebox{.48\textwidth}{!}{\includegraphics*{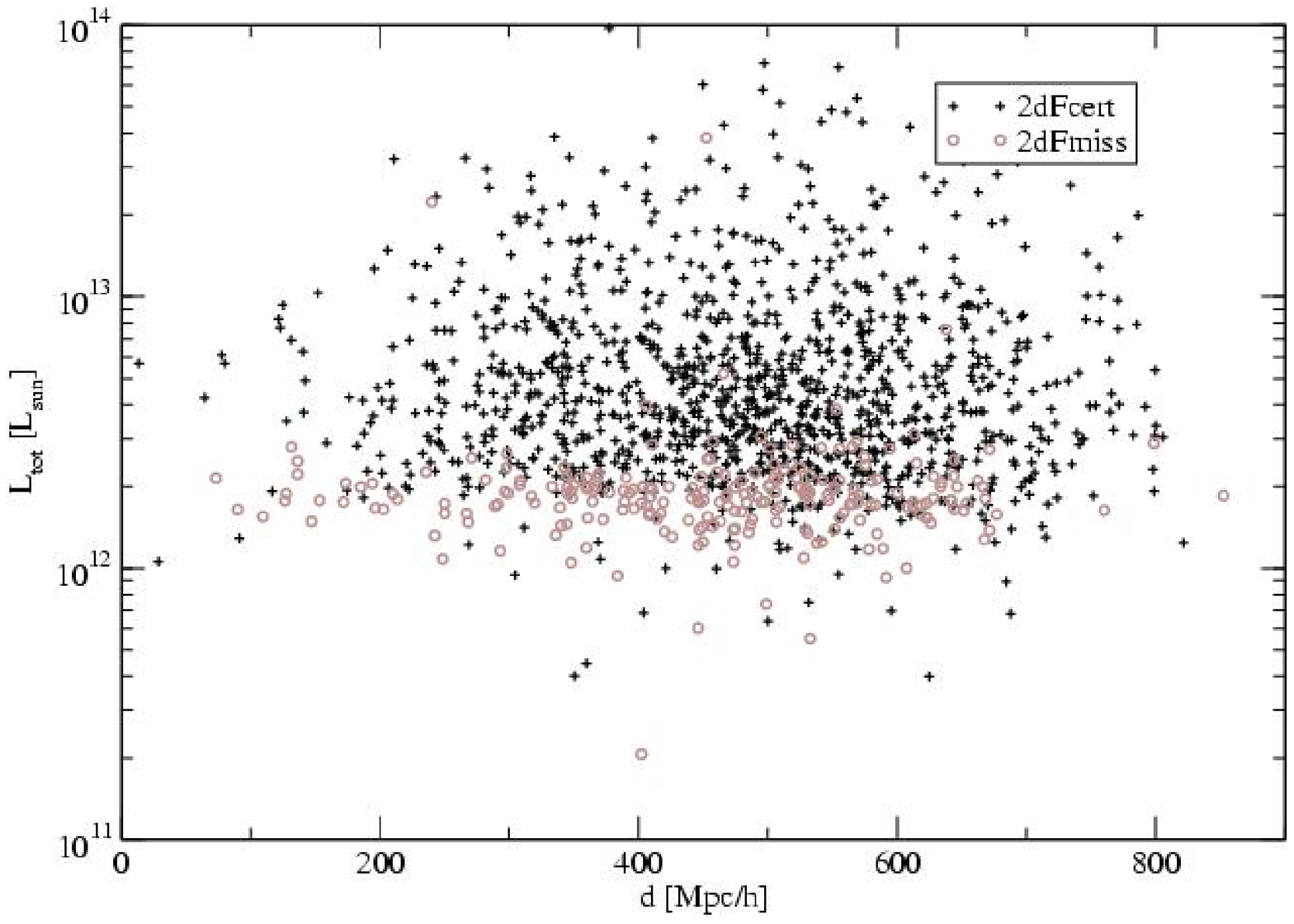}}
\\
\caption{Left panel: the number of galaxies in superclusters of the
  sample Mill.F8 as a function of distance from the observer.  The
  right panel shows luminosities of superclusters of the sample
  Mill.F8 as a function of distance (crosses).  Open gray circles show
  luminosities of superclusters of the total sample Mill.A8 which have
  no counterpart in the sample Mill.F8 (missing superclusters).  
}
\label{fig:11}
\end{figure*}

This analysis shows that there are good reasons to consider highest peaks of
the smoothed density field as centres of superclusters.

\subsection{Analysis of simulated flux-limited samples}

We have constructed  simulated flux- and volume-limited subsamples of
Millennium Simulation galaxies.  Using these subsamples we found
superclusters and derived their properties for two cases.  First, we
used the density field of all galaxies, and calculated total
luminosities of superclusters for two cases, using full data (Mill.A8
sample) and the simulated 2dF sample Mill.F8. In the second case
estimated total luminosities of galaxies were found as for 2dFGRS
applying Eq.~(\ref{eq:weight}).  The luminosity function was taken
directly from the simulation data, as shown in Fig.~\ref{fig:5}.  This
case allows us to estimate the  errors of estimated total
luminosities of superclusters using restored galaxy total
luminosities. Here the lists of superclusters contain identical
entries, only the number of galaxies within them and the estimated
total luminosities differ. In this case we do not take into account
the fact that the density field also has errors due to the use of
incomplete (flux-limited) data.

To get an idea of the scale of the external errors of calculated
total luminosities we  calculated the smoothed density field using 
galaxies of the subsample Mill.F8.  In this case the identification of
superclusters with the first sample is more difficult, since centre
coordinates may differ.  For identification we identified  for every
supercluster of the subsample Mill.F8 (and Mill.V8) the closest system
among the sample Mill.A8. 

The left panel of Fig.~\ref{fig:10} shows the fraction of the number of
galaxies in superclusters of subsamples Mill.F8 and Mill.V8 with respect
to the respective number in the sample Mill.A8.  We see that at small
distances this fraction for the subsample Mill.F8 is close to unity,
i.e. almost all galaxies are present also in the flux-limited
subsample. With increasing distance the fraction gradually
decreases. The volume-limited subsample has at large distances from the
observer  a behaviour similar to the full flux-limited subsample,
but on distances less than 400~\Mpc\ the fraction remains constant at
a level about 0.2. Some data-points above the main ridge at large
distance are due to misidentification of superclusters in our
automated procedure. 

In the right panel of Fig.~\ref{fig:10} we plot the relative error of
the total luminosity of superclusters as a function of the distance
from the observer. Black symbols show internal errors
(i.e. supercluster volumes were found using identical density
fields, and expected total luminosities were found for complete and
simulated 2dF data), gray symbols show external errors (superclusters of the
subsample Mill.F8 were found using the density field determined by the
same subsample of simulated 2dF galaxies).  We see that at large
distance from the observer ($d> 500$~\Mpc) both internal and external
errors become large, since the number of galaxies in superclusters
becomes too small. At intermediate distances $200 \leq d \leq
400$~\Mpc\ internal errors are surprisingly small, and there are
practically no systematic errors. External errors are larger, and have
a negative tail, i.e. luminosities of superclusters determined from
incomplete (flux-limited) data are systematically lower than those
calculated using full data. Partly this difference is due to the fact
that some superclusters of the sample Mill.A8 are split into smaller
systems in the subsample Mill.F8.

Fig.~\ref{fig:11} (left panel) shows the number of galaxies in
superclusters of samples Mill.A8 and Mill.F8 as a function of
distance.  We see that the true number of galaxies in superclusters
exceeds 200 with only a few exceptions (remember that the galaxy
sample of the Millennium Simulation is complete for luminosities
exceeding an absolute magnitude $-17.4$ in the {\tt r}-band).  In
superclusters identified using flux-limited galaxy samples the number
of galaxies decreases with distance.  This decrease follows the same
law as the fraction shown in Fig.~\ref{fig:10} for simulated 2dFGRS
superclusters.

The comparison of lists of superclusters of samples Mill.A8 and
Mill.F8 shows that about 200 superclusters of the full sample Mill.A8
have no counterparts in the sample Mill.F8, based on the flux-limited
sample of galaxies. In other words, these superclusters are too weak
to meet our selection criterion. We show the distribution of
luminosities of missing superclusters as a function of distance in the
right panel of Fig.~\ref{fig:11} by gray symbols. For comparison
luminosities of all detected superclusters are also shown.  We see
that all missing superclusters have  low luminosities. In other words,
the luminosity function of superclusters found on the basis of
flux-limited galaxy samples is biased and needs to be corrected in the
range of poor superclusters.

\section{2dF supercluster catalogue}

Based on the experience of the study of simulated superclusters we
shall use for the compilation of the final catalogue of 2dFGRS
superclusters the DF method. To calculate the luminosity density field
we use an Epanechnikov kernel with radius 8~\Mpc, and a rectangular grid
of cell size 1~\Mpc.  To minimize the size of the density field box we treat
Northern and Southern regions of 2dF separately.  The coordinate
system was rotated along the vertical axis so that the sample starts
at $x-$axis: 
\begin{equation}
x = d \cos(\delta)\cos(\alpha - \alpha_0),\\
\end{equation}
\begin{equation}
y = d \cos(\delta)\sin(\alpha - \alpha_0),\\
\end{equation}
\begin{equation}
z = d \sin(\delta),
\label{eq:rot}
\end{equation}
where $\alpha_0$ is the minimal value of the Right Ascension for the
sample, which is $148.13^\circ$ and $-34.40^\circ$ for the
Northern and Southern samples, respectively. After the rotation of
coordinates around the $z$-axis both Northern and Southern samples fit
in the first quadrant, and $x, y$ coordinates are non-negative. The
size of the box along the vertical axis is determined by extreme
$z-$coordinates of galaxies within the observed regions.  Densities
were calculated using the total estimated luminosities of galaxies,
and then reduced to the mean density over the whole sample. To avoid
the inclusion of unobserved regions all cells outside the
observational window were marked.

The next step in the selection of superclusters is the proper choice
of the threshold density $D_0$ to separate high and low-density galaxy
systems.  We compiled supercluster catalogues in a wide range of
threshold densities from 1 to 7 in units of the mean luminosity density
of the sample.  For each threshold density value we found the number
of superclusters, $N_{scl}$, and calculated the maximal diameter of
the largest system found, $L_{scl}$.  Fig.~\ref{fig:03} shows results
of these calculations.  Based on these data we apply threshold density
4.6 for our final supercluster catalogue.  Using this threshold
density a few superclusters still have diagonal sizes exceeding
120~\Mpc; these superclusters split into subsystems when a larger
threshold density is used.

In our preliminary analysis we used a minimal volume 200~(\Mpc)$^3$.
The analysis shows that this limit is too high and excludes a number
of small superclusters of the Local Supercluster class. Thus in our
final catalogue we have used a smaller limiting volume of 100~(\Mpc)$^3$,
which corresponds to limiting radius 2.8~\Mpc. Using this limit we
include practically all galaxy systems which exceed the chosen
threshold limit into our supercluster list, and exclude only systems
which have a very small fraction of their volume above the threshold.
At this level noise due to random errors of corrected galaxy
luminosities becomes large.  Remember that the use of smoothing with
8~\Mpc\ radius means that all galaxies and groups within this radius
are used in the calculation of the density field; thus even the
smallest superclusters represent galaxy samples located in a much
larger volume than the volume above the density threshold.

The lists of all 2dF groups and single galaxies were searched to find
members of superclusters. The number of galaxies in superclusters as a
function of the distance from the observer is shown in
Fig.~\ref{fig:04}.  As expected, this number decreases with distance.
In very poor and distant superclusters the number of galaxies detected may
fall below 3.  These very poor superclusters have been excluded from
our supercluster list.  Also, as our analysis of simulated
superclusters has shown, some parameters of very distant superclusters
have rather large statistical uncertainties. For this reason we have
divided our supercluster lists into two parts: the main sample
(denoted A) contains superclusters up to distance 520~\Mpc, and the
supplementary sample (denoted B) has more distant superclusters.

We note that only about 1/3 of all galaxies of the 2dF survey are
members of superclusters. The remaining galaxies and groups also
belong to galaxy systems, but these systems are weaker and form in the
density field enhancements with peak density less than our adopted
threshold value 4.6.

We also compiled lists of compact high-density peaks of the density
field -- DF-clusters, using the low-resolution density field and
threshold density 5.0 (in units of the mean density). DF-clusters are
some equivalent to rich Abell-type clusters.  Since luminosities were
corrected to take into account galaxies outside the visibility window,
DF-clusters form a volume-limited sample.  DF-clusters are useful in
the further identification of rich clusters of galaxies of the Abell
cluster class. The right panel of Fig.\ref{fig:04} shows the distance
dependence of the number of DF-clusters in superclusters.

We calculated the luminous mass around the center of DF-cluster in a
box containing the parent cell and all surrounding cells (altogether
27 cells). The most luminous group (from the list by T06) in this
box was considered as the main group/cluster of the supercluster, and the
brightest galaxy of the main cluster was taken as the main galaxy of
the supercluster. The center  of the supercluster was identified with
the center of the main galaxy. Some poor superclusters do not contain
peaks above the threshold 5.0 used in the search of DF-clusters. In
these cases the most luminous group was considered as the center of
the supercluster.

To have an idea of the spatial distribution of luminous matter we can
have a look at respective density fields den-Ngr-570-86-ep8.fits and
den-Sgr-516-312-ep8.fits on our web-page. These fields can be seen in
$x,y,z$ coordinates using the viewer ds9. We see the multi-nucleus
character of most superclusters.  Note also the asymmetry of the
distribution of galaxies in superclusters.

The final catalogue of 2dFGRS superclusters consists of four lists,
two for each Galaxy hemisphere, the main lists A for superclusters up
to distance 520~\Mpc\ and supplementary lists B for more distant
systems. These lists are given in the electronic supplement of the
paper.  The lists are ordered according to increasing RA, separately
for the Northern and Southern hemispheres, but a common id-numeration
for lists A and B.  The lists have the following entries.

\begin{enumerate}

\item{} Identification number.
\item{} Equatorial coordinates (for the epoch 2000).
\item{} The distance  $d$.
\item{} The minimal size of the supercluster $D_{min} =
  min(dx,dy,dz)$, where $dx$, $dy$, and $dz$ are sizes of the
  supercluster along coordinates $x,~y,~z$; the sizes are determined
  from extreme coordinates of the density field above threshold
  density along coordinate axes.
\item{} The maximal diameter (diagonal of the box containing the
  supercluster) $D_{max} = (dx^2 + dy^2 + dz^2)^{1/2}$.
\item{} The effective diameter   $D_e$ (the diameter of the sphere,
  equal to the volume of the supercluster).
\item{} The ratio of the mean to effective diameter:
$\epsilon_0 = D_m/D_e$, here $D_m = D_{max}/3^{1/3}$ is the mean diameter.
This parameter characterizes
the compactness of the system.
\item{} The center offset parameter, $\Delta_o = ((x_0 - x_m)^2 +
(y_0 - y_m)^2 + (z_0 - z_m)^2)^{1/2}$; here $x_0, y_0, z_0$ are
coordinates of the geometric center of the supercluster, found on the
basis of extreme values of coordinates, and $x_m, y_m, z_m$ are
coordinates of the dynamical center (main cluster) of the
supercluster. This parameter characterises the asymmetry of the
supercluster.
\item{} The peak density,  $\delta_p$ (in units of the mean density).
\item{} The mean density,  $\delta_m$, (in units of the mean density).
\item{} The number of galaxies in the supercluster, $N_{gal}$.
\item{} The number of groups, $N_{gr}$ (including groups with only
  one visible   galaxy) in the supercluster.
\item{} The multiplicity of the supercluster, $N_{cl}$ (the number of
  DF-clusters).
\item{} The identification of the main cluster according to the
  group catalogue by T06.
\item{} The number of visible galaxies in the main cluster.
\item{} The luminosity of the main galaxy, $L_m$.
\item{} The total estimated luminosity of the supercluster,
$L_{tot}$, in photometric system ${\rm b_j}$, expressed in Solar
units. The total luminosity was found by summing all estimated total
luminosities of member groups of the supercluster.

\end{enumerate}

The lists of superclusters of the main catalogue, having total
estimated luminosities above $2.5\times 10^{12}$ $L_\odot/h^2$, are
given in Tables~\ref{tab:Nscl} and \ref{tab:Sscl}.  
The full lists of superclusters, both the main and
supplementary, are available at the web-site of Tartu Observatory
\texttt{http://www.aai.ee/$\sim$maret/2dfscl.html}.  Density fields
in fits format are also available for 2dFGRS and Millennium Simulation
samples, see readme.txt for details.

\section{Conclusions}

In this paper we have used the 2dF Galaxy Redshift Survey to compose a new
catalogue of superclusters of galaxies.  Our main conclusions are the
following. 

\begin{itemize}

\item{} To analyse selection effects and possible biases, and to find
  suitable parameters to select superclusters of galaxies, we analysed
  simulated superclusters found using the Millennium Simulation of the
  evolution of the Universe. 

\item{} We calculated the density field using the 2dF Galaxy Redshift
  Survey in the Northern and Southern region, applying smoothing with
  an Epanechnikov kernel of radius 8 \Mpc, and using weights for
  galaxies which take allowance for faint galaxies outside the
  observational window of apparent magnitudes.
  
\item{} Using the smoothed density field we identified superclusters
  of galaxies as galaxy systems which occupy regions in the density
  field above the threshold density 4.6 in units of the mean density,
  and having a minimal volume of 100~(\Mpc)$^3$, separately for the
  Northern and Southern regions of the 2dFGRS.

\item{} We calculated for all superclusters their main parameters:
  equatorial coordinates, distances, minimal, maximal and effective
  diameters, the number of galaxies, groups and DF-clusters,
  luminosities of main clusters and their main galaxies, total
  luminosities, overdensities and separation between geometrical and 
  mass center.

\item{} The analysis of the properties of superclusters shows that our
  supercluster samples are free from known biases.

\end{itemize}

\begin{acknowledgements}
  
   We are pleased to thank the 2dF GRS Team for the publicly available
  final data release. The Millennium Simulation used in this paper was
  carried out by the Virgo Supercomputing Consortium at the Computing
  Centre of the Max-Planck Society in Garching. This research has made
  use of SAOImage DS9, developed by Smithsonian Astrophysical
  Observatory. The semi-analytic galaxy catalogue is publicly
  available at http://www.mpa-garching.mpg.de/galform/agnpaper.  The
  present study was supported by Estonian Science Foundation grants
  No.  4695, 5347 and 6104 and 6106, and Estonian Ministry for Education and
  Science support by grant TO 0060058S98. This work has also been
  supported by the University of Valencia through a visiting
  professorship for Enn Saar and by the Spanish MCyT project
  AYA2003-08739-C02-01.  J.E.  thanks Astrophysikalisches Institut
  Potsdam (using DFG-grant 436 EST 17/2/05) for hospitality where part
  of this study was performed.

\end{acknowledgements}

{
\scriptsize
\vskip-2.5cm 
\begin{table*}[hp] 
\begin{center} 
\caption{The list of rich 2dF Northern superclusters} 
\begin{tabular}{rrrrrrrrrrrrrccc} 
\\ 
\hline 
\\ 
Id & RA & DEC & $d$ & $D_{min}$ & $D_{max}$ & $D_0$ &$\epsilon_0$& $\Delta_o$ & $\delta_p$& $\delta_m$&
$N_{gr}$& $N_{cl}$ &  $L_m$  & $L_{tot}$ \\
   & deg& deg & Mpc & Mpc & Mpc   & Mpc &   & Mpc  & & & & & & \\
\\ 
\hline 
\\ 
   5 & 149.58 &  -4.64 & 457.6 &  19.0 &  45.2 &  18.6 &   1.4 &   3.3 &  10.9 &   6.6 &   102 &     5 &  0.3553E+11 &  0.5460E+13 \\  
   9 & 150.47 &  -0.86 & 398.8 &  19.0 &  37.1 &  15.0 &   1.4 &   9.2 &   5.7 &   5.6 &    52 &     3 &  0.6251E+11 &  0.2675E+13 \\  
  13 & 152.01 &   0.57 & 288.1 &  31.0 &  89.7 &  27.5 &   1.9 &  25.3 &  12.9 &   6.9 &  1145 &    10 &  0.4872E+11 &  0.1646E+14 \\  
  17 & 153.54 &  -4.22 & 467.4 &  20.0 &  66.2 &  21.5 &   1.8 &   8.9 &   9.3 &   6.0 &   120 &     8 &  0.4714E+11 &  0.8309E+13 \\  
  20 & 155.11 &  -2.54 & 184.5 &  17.0 &  40.0 &  15.5 &   1.5 &   8.2 &   7.2 &   5.7 &   556 &     2 &  0.3929E+11 &  0.3188E+13 \\  
  27 & 156.91 &   1.86 & 440.4 &  15.0 &  33.5 &  15.4 &   1.3 &   8.2 &   6.4 &   6.2 &    68 &     4 &  0.1652E+11 &  0.3327E+13 \\  
  37 & 160.34 &  -5.90 & 384.5 &  18.0 &  67.0 &  22.2 &   1.7 &  28.2 &  12.0 &   6.7 &   359 &     9 &  0.3478E+11 &  0.9555E+13 \\  
  38 & 160.57 &  -3.74 & 509.6 &  23.0 &  41.6 &  14.6 &   1.6 &   6.8 &  10.5 &   6.3 &    34 &     4 &  0.3574E+11 &  0.3080E+13 \\  
  76 & 170.64 &   0.45 & 302.5 &  17.0 &  42.7 &  19.3 &   1.3 &   8.2 &   6.7 &   6.7 &   420 &     5 &  0.4723E+11 &  0.5933E+13 \\  
  77 & 170.77 &   1.03 & 220.4 &  10.0 &  30.7 &  13.9 &   1.3 &   7.7 &   8.3 &   6.2 &   315 &     2 &  0.4605E+11 &  0.2527E+13 \\  
  78 & 170.87 &   0.32 & 425.0 &  12.0 &  42.4 &  15.0 &   1.6 &  11.8 &   5.6 &   5.5 &    57 &     5 &  0.5354E+11 &  0.2504E+13 \\  
  82 & 172.65 &   1.46 & 370.2 &  18.0 &  38.8 &  17.6 &   1.3 &  10.9 &   6.7 &   6.1 &   187 &     3 &  0.1872E+11 &  0.4419E+13 \\  
  92 & 175.90 &  -1.73 & 313.3 &  24.0 &  52.4 &  19.0 &   1.6 &   7.4 &  11.1 &   6.5 &   315 &     3 &  0.2367E+11 &  0.5366E+13 \\  
  97 & 176.85 &  -2.85 & 359.6 &  17.0 &  36.7 &  16.3 &   1.3 &   9.8 &   7.6 &   5.9 &   129 &     5 &  0.8599E+11 &  0.3051E+13 \\  
  99 & 177.62 &  -0.60 & 399.3 &  40.0 &  76.6 &  26.2 &   1.7 &  15.0 &   8.4 &   6.2 &   472 &    13 &  0.6194E+11 &  0.1421E+14 \\  
 101 & 178.42 &  -2.29 & 510.7 &  13.0 &  35.1 &  16.0 &   1.3 &   6.6 &   7.4 &   6.1 &    29 &     5 &  0.3994E+11 &  0.2841E+13 \\  
 108 & 180.44 &  -0.20 & 481.7 &  43.0 &  79.8 &  26.6 &   1.7 &  18.4 &   8.7 &   6.3 &   169 &    24 &  0.9918E+11 &  0.1463E+14 \\  
 118 & 183.28 &  -3.72 & 489.4 &  12.0 &  28.5 &  14.6 &   1.1 &   5.7 &   8.2 &   6.7 &    38 &     1 &  0.3733E+11 &  0.2828E+13 \\  
 120 & 183.61 &  -3.57 & 512.4 &  42.0 &  96.0 &  30.8 &   1.8 &  24.9 &  12.5 &   7.6 &   207 &    19 &  0.4793E+11 &  0.2449E+14 \\  
 127 & 185.45 &   0.34 & 458.4 &  37.0 &  76.8 &  24.9 &   1.8 &  13.8 &  10.1 &   7.1 &   176 &    10 &  0.7087E+11 &  0.1364E+14 \\  
 136 & 190.07 &  -4.44 & 395.5 &  20.0 &  50.0 &  20.4 &   1.4 &  17.4 &  10.5 &   6.8 &   251 &     2 &  0.5639E+11 &  0.7551E+13 \\  
 137 & 190.09 &  -2.56 & 486.5 &  17.0 &  42.5 &  18.2 &   1.4 &  10.7 &  15.4 &   7.3 &    73 &     3 &  0.8384E+11 &  0.5971E+13 \\  
 140 & 191.19 &  -1.08 & 430.1 &  19.0 &  36.6 &  14.8 &   1.4 &   9.3 &   6.2 &   5.4 &    62 &     4 &  0.9454E+11 &  0.2733E+13 \\  
 147 & 193.74 &  -2.30 & 500.2 &  20.0 &  50.8 &  21.4 &   1.4 &   7.2 &  12.5 &   6.7 &    82 &     4 &  0.5705E+11 &  0.7776E+13 \\  
 152 & 194.71 &  -1.74 & 251.1 &  36.0 & 112.7 &  35.7 &   1.8 &  14.8 &  14.4 &   7.7 &  3591 &    18 &  0.5353E+11 &  0.3783E+14 \\  
 155 & 196.07 &   1.35 & 511.9 &  20.0 &  38.2 &  17.2 &   1.3 &   5.1 &  10.9 &   6.6 &    43 &     4 &  0.2870E+11 &  0.4448E+13 \\  
 162 & 198.32 &  -2.18 & 418.9 &  32.0 &  75.4 &  20.3 &   2.2 &  10.6 &   5.3 &   5.9 &   196 &     7 &  0.1395E+11 &  0.7419E+13 \\  
 170 & 200.94 &   1.08 & 320.4 &  30.0 &  56.8 &  21.5 &   1.5 &  14.5 &   9.1 &   6.7 &   415 &     8 &  0.2950E+11 &  0.8103E+13 \\  
 181 & 203.14 &  -3.08 & 508.7 &  41.0 &  83.2 &  31.2 &   1.5 &  12.8 &  18.3 &   8.0 &   200 &    15 &  0.7200E+11 &  0.2628E+14 \\  
 193 & 208.59 &  -1.05 & 432.1 &  28.0 &  61.9 &  23.1 &   1.5 &   9.5 &  11.7 &   6.7 &   197 &     8 &  0.6642E+11 &  0.9617E+13 \\  
 196 & 209.56 &   1.18 & 480.7 &  19.0 &  40.9 &  18.1 &   1.3 &   8.7 &   5.7 &   6.8 &    69 &     4 &  0.4161E+11 &  0.4950E+13 \\  
 205 & 213.68 &  -0.39 & 405.3 &  28.0 &  63.7 &  23.8 &   1.5 &  14.3 &  28.9 &   9.3 &   215 &     5 &  0.1701E+12 &  0.1309E+14 \\  
 210 & 216.18 &  -2.00 & 506.1 &  15.0 &  28.4 &  14.6 &   1.1 &   3.3 &  10.5 &   7.3 &    15 &     1 &  0.5666E+11 &  0.2781E+13 \\  
 220 & 219.41 &  -0.29 & 399.9 &  38.0 &  90.7 &  28.5 &   1.8 &  29.2 &   9.8 &   6.3 &   426 &    16 &  0.6000E+11 &  0.1812E+14 \\  
 225 & 220.83 &  -0.67 & 436.0 &  15.0 &  28.4 &  14.9 &   1.1 &   3.6 &  10.0 &   6.7 &    55 &     1 &  0.3814E+11 &  0.2756E+13 \\  
\\
\hline
\label{tab:Nscl}
\end{tabular}
\end{center}
Distance and sizes are given in \Mpc.
\end{table*}
}
{
\scriptsize
\vskip-2.5cm 
\begin{table*}[hp] 
\begin{center} 
\caption{The list of rich 2dF Southern superclusters} 
\begin{tabular}{rrrrrrrrrcrccccc} 
\\ 
\hline 
\\ 
Id & RA & DEC & $d$ & $D_{min}$ & $D_{max}$ & $D_0$ &$\epsilon_0$& $\Delta_o$ & $\delta_p$& $\delta_m$&
$N_{gr}$& $N_{cl}$ &  $L_m$  & $L_{tot}$ \\
   & deg& deg & Mpc & Mpc & Mpc   & Mpc &   & Mpc  & & & & & & \\
\\ 
\hline 
\\ 
   5 &   1.85 & -28.06 & 177.4 &  20.0 &  45.7 &  19.7 &   1.3 &   6.8 &   7.5 &   6.2 &   952 &     5 &  0.4755E+11 &  0.4824E+13 \\  
  10 &   3.02 & -27.42 & 362.6 &  39.0 &  96.3 &  25.4 &   2.2 &  15.5 &   5.3 &   5.9 &   535 &    17 &  0.4525E+11 &  0.1155E+14 \\  
  11 &   3.49 & -27.10 & 436.8 &  17.0 &  38.8 &  17.3 &   1.3 &   6.7 &  10.0 &   6.4 &   101 &     3 &  0.4848E+11 &  0.3902E+13 \\  
  18 &   5.15 & -33.91 & 459.5 &  16.0 &  40.5 &  16.5 &   1.4 &   8.5 &   8.2 &   6.1 &    75 &     3 &  0.2783E+11 &  0.3581E+13 \\  
  19 &   5.17 & -25.73 & 414.7 &  18.0 &  33.0 &  17.2 &   1.1 &   2.9 &  10.7 &   6.8 &    91 &     2 &  0.1359E+12 &  0.3566E+13 \\  
  34 &   9.86 & -28.94 & 326.3 &  60.0 & 140.1 &  40.9 &   2.0 &  21.6 &  16.9 &   8.1 &  3175 &    24 &  0.4528E+11 &  0.4975E+14 \\  
  51 &  13.98 & -30.08 & 455.4 &  33.0 &  77.2 &  25.6 &   1.7 &  13.6 &  14.9 &   7.6 &   272 &     7 &  0.4022E+11 &  0.1278E+14 \\  
  60 &  16.54 & -26.29 & 376.5 &  17.0 &  41.1 &  15.7 &   1.5 &   9.5 &   7.9 &   5.7 &   132 &     4 &  0.2102E+11 &  0.2905E+13 \\  
  64 &  17.74 & -33.10 & 475.4 &  12.0 &  34.1 &  15.6 &   1.3 &   6.9 &   8.3 &   6.2 &    50 &     4 &  0.9654E+11 &  0.2950E+13 \\  
  78 &  21.00 & -33.35 & 513.6 &  34.0 &  93.1 &  29.7 &   1.8 &  19.2 &  17.6 &   6.8 &   254 &    20 &  0.3307E+11 &  0.1697E+14 \\  
  84 &  21.97 & -34.16 & 383.9 &  21.0 &  50.6 &  20.0 &   1.5 &  10.6 &  10.5 &   7.5 &   225 &     3 &  0.4226E+11 &  0.5491E+13 \\  
  87 &  23.48 & -27.53 & 362.0 &  17.0 &  49.2 &  17.8 &   1.6 &  15.7 &   9.2 &   6.2 &   166 &     4 &  0.4329E+11 &  0.3867E+13 \\  
  88 &  23.57 & -26.11 & 460.3 &  24.0 &  50.7 &  20.0 &   1.5 &   8.4 &  11.9 &   6.9 &   105 &     2 &  0.1200E+12 &  0.5457E+13 \\  
  94 &  25.44 & -30.60 & 483.7 &  38.0 &  75.3 &  27.0 &   1.6 &   4.6 &  10.0 &   6.6 &   245 &    15 &  0.6197E+11 &  0.1316E+14 \\  
  97 &  27.22 & -31.42 & 432.5 &  27.0 &  53.7 &  21.5 &   1.4 &   7.2 &   8.9 &   6.5 &   194 &     5 &  0.2499E+11 &  0.7256E+13 \\  
 109 &  30.94 & -26.79 & 329.1 &  15.0 &  39.0 &  17.9 &   1.3 &   1.9 &   6.1 &   7.1 &   249 &     2 &  0.2811E+11 &  0.4286E+13 \\  
 112 &  31.38 & -34.62 & 475.3 &  27.0 &  60.9 &  22.7 &   1.5 &  12.4 &  10.6 &   6.8 &   148 &     7 &  0.3140E+11 &  0.8057E+13 \\  
 115 &  32.38 & -28.85 & 391.6 &  28.0 &  59.6 &  21.0 &   1.6 &  14.4 &   5.5 &   5.8 &   265 &    10 &  0.1617E+11 &  0.6458E+13 \\  
 116 &  32.60 & -33.01 & 357.3 &  17.0 &  47.8 &  19.1 &   1.4 &  11.1 &   6.6 &   6.3 &   230 &     3 &  0.4483E+11 &  0.4801E+13 \\  
 126 &  34.36 & -29.43 & 314.6 &  17.0 &  40.8 &  17.7 &   1.3 &  12.7 &   7.9 &   6.3 &   291 &     3 &  0.4460E+11 &  0.3683E+13 \\  
 130 &  35.03 & -28.77 & 484.5 &  27.0 &  86.0 &  30.1 &   1.6 &  13.1 &   7.9 &   7.0 &   292 &    19 &  0.3013E+11 &  0.1884E+14 \\  
 148 &  41.09 & -26.27 & 386.3 &  31.0 &  63.5 &  23.3 &   1.6 &   5.2 &  11.6 &   6.9 &   328 &    11 &  0.2811E+11 &  0.8765E+13 \\  
 149 &  41.41 & -34.25 & 506.2 &  16.0 &  34.0 &  16.6 &   1.2 &   5.5 &  10.0 &   6.7 &    67 &     3 &  0.2502E+11 &  0.3516E+13 \\  
 152 &  42.21 & -26.00 & 306.0 &  16.0 &  38.2 &  15.6 &   1.4 &   7.4 &   6.5 &   5.5 &   180 &     4 &  0.2678E+11 &  0.2741E+13 \\  
 153 &  42.57 & -26.42 & 461.3 &  24.0 &  44.5 &  16.8 &   1.5 &  12.2 &   9.1 &   5.8 &    64 &     8 &  0.4152E+11 &  0.3518E+13 \\  
 161 &  46.01 & -31.64 & 510.5 &  22.0 &  42.9 &  17.8 &   1.4 &  11.7 &   9.1 &   6.3 &    50 &     3 &  0.3963E+11 &  0.4215E+13 \\  
 167 &  47.93 & -26.94 & 198.7 &  24.0 &  43.9 &  19.9 &   1.3 &   6.2 &  14.2 &   7.2 &   771 &     2 &  0.4938E+11 &  0.5266E+13 \\  
 179 &  52.32 & -30.10 & 508.4 &  14.0 &  49.5 &  17.7 &   1.6 &   5.3 &   5.1 &   5.7 &    58 &     8 &  0.2036E+11 &  0.4405E+13 \\  
 180 &  52.33 & -26.53 & 421.1 &  44.0 &  90.6 &  25.0 &   2.1 &  25.0 &   9.4 &   6.2 &   263 &    10 &  0.4963E+11 &  0.1021E+14 \\  
 184 &  53.82 & -28.68 & 301.7 &  13.0 &  29.4 &  15.2 &   1.1 &   6.2 &   7.2 &   6.5 &   156 &     1 &  0.2297E+11 &  0.2512E+13 \\  
 185 &  53.93 & -29.63 & 393.8 &  19.0 &  51.6 &  18.6 &   1.6 &  18.8 &   4.8 &   5.6 &   151 &     7 &  0.3211E+11 &  0.4318E+13 \\  
 190 & 327.23 & -30.67 & 352.2 &  19.0 &  43.0 &  15.1 &   1.6 &  11.7 &   9.1 &   6.0 &   122 &     5 &  0.2405E+11 &  0.2583E+13 \\  
 200 & 330.81 & -24.36 & 466.5 &  30.0 &  61.1 &  24.6 &   1.4 &   2.7 &  20.4 &   8.7 &   155 &     6 &  0.4344E+11 &  0.1242E+14 \\  
 204 & 331.43 & -27.84 & 268.2 &  16.0 &  44.0 &  18.0 &   1.4 &   5.6 &   7.0 &   5.8 &   342 &     5 &  0.6541E+11 &  0.4041E+13 \\  
 205 & 331.45 & -25.20 & 515.4 &  13.0 &  38.4 &  16.4 &   1.4 &   5.9 &   8.4 &   6.0 &    23 &     3 &  0.6376E+11 &  0.3097E+13 \\  
 209 & 332.72 & -29.87 & 465.8 &  20.0 &  45.8 &  20.2 &   1.3 &   6.5 &   9.7 &   6.8 &    91 &     5 &  0.3213E+11 &  0.5512E+13 \\  
 217 & 334.75 & -34.76 & 449.1 &  66.0 & 126.4 &  39.8 &   1.8 &  23.4 &  11.2 &   6.9 &   938 &    42 &  0.3692E+11 &  0.4320E+14 \\  
 220 & 335.53 & -31.32 & 343.9 &  15.0 &  36.8 &  16.6 &   1.3 &   8.0 &   9.1 &   6.2 &   169 &     6 &  0.2286E+11 &  0.3292E+13 \\  
 221 & 335.77 & -29.33 & 506.6 &  20.0 &  43.8 &  16.8 &   1.5 &   6.5 &   7.0 &   6.1 &    37 &     7 &  0.2982E+11 &  0.3324E+13 \\  
 222 & 336.90 & -30.58 & 169.0 &  15.0 &  35.2 &  15.8 &   1.3 &   7.1 &   9.8 &   6.2 &   473 &     2 &  0.3426E+11 &  0.2660E+13 \\  
 229 & 341.30 & -31.91 & 505.2 &  13.0 &  33.1 &  15.8 &   1.2 &   3.3 &   9.8 &   6.6 &    41 &     2 &  0.2632E+11 &  0.2877E+13 \\  
 240 & 343.16 & -26.04 & 441.9 &  26.0 &  52.4 &  21.5 &   1.4 &   7.9 &  13.4 &   6.8 &   171 &     6 &  0.4852E+11 &  0.6829E+13 \\  
 247 & 345.22 & -31.62 & 514.6 &  13.0 &  29.9 &  15.4 &   1.1 &   3.2 &   6.3 &   6.8 &    28 &     2 &  0.2949E+11 &  0.2653E+13 \\  
 253 & 346.13 & -32.51 & 245.9 &  29.0 &  68.8 &  19.6 &   2.0 &  11.2 &   9.8 &   6.2 &   521 &     7 &  0.2588E+11 &  0.5180E+13 \\  
 260 & 347.89 & -29.08 & 337.5 &  18.0 &  39.8 &  18.5 &   1.2 &   5.6 &   8.9 &   6.3 &   220 &     4 &  0.2053E+11 &  0.4487E+13 \\  
 267 & 348.85 & -24.69 & 421.4 &  25.0 &  53.6 &  21.6 &   1.4 &  11.4 &   8.2 &   6.7 &   173 &     6 &  0.1890E+11 &  0.7051E+13 \\  
 270 & 350.71 & -28.24 & 506.4 &  21.0 &  51.2 &  17.2 &   1.7 &   7.6 &   8.8 &   6.1 &    37 &     8 &  0.3358E+11 &  0.3662E+13 \\  
 271 & 351.13 & -29.93 & 456.4 &  27.0 &  76.0 &  26.4 &   1.7 &   5.9 &  13.7 &   7.1 &   253 &     8 &  0.2550E+11 &  0.1349E+14 \\  
 276 & 352.36 & -30.15 & 306.9 &  29.0 &  61.8 &  20.4 &   1.8 &  18.8 &   8.2 &   5.8 &   371 &     8 &  0.1736E+11 &  0.5993E+13 \\  
 282 & 353.01 & -34.42 & 397.2 &  35.0 &  66.6 &  22.2 &   1.7 &  22.0 &  11.3 &   6.5 &   255 &    14 &  0.9439E+11 &  0.8168E+13 \\  
 303 & 357.53 & -26.99 & 452.1 &  22.0 &  41.0 &  16.7 &   1.4 &  10.1 &   9.5 &   6.1 &    71 &     5 &  0.3256E+11 &  0.3168E+13 \\  
 313 & 359.87 & -26.13 & 398.4 &  18.0 &  32.9 &  16.0 &   1.2 &   5.2 &   8.4 &   6.0 &    85 &     4 &  0.4653E+11 &  0.2611E+13 \\  
\\
\hline
\label{tab:Sscl}
\end{tabular}
\end{center}
\end{table*}
}

\end{document}